\documentclass[a4paper]{aa}
\usepackage{graphicx}
\usepackage{txfonts}
\usepackage{natbib}
\usepackage{aalongtable}

\bibpunct{(}{)}{;}{a}{}{,}

\newcommand{\kms}{km~s$^{-1}$\,}

\newcommand {\sm}{$M_{\odot}$\,}
\newcommand {\vi}{$V-I$}

\begin{document}

\title{The DART imaging and CaT survey of the Fornax Dwarf
Spheroidal Galaxy}

\author{G. Battaglia\inst{1} \and 
E. Tolstoy\inst{1} \and 
A. Helmi\inst{1} \and  M.J. Irwin\inst{2} \and 
B. Letarte\inst{1} \and  P. Jablonka\inst{3} \and 
V. Hill\inst{4} \and 
K.A. Venn\inst{5} \and  M.D. Shetrone\inst{6} \and 
N. Arimoto\inst{7} \and 
F. Primas\inst{8} \and 
A. Kaufer\inst{9} \and  P. Francois\inst{4} \and T. Szeifert\inst{9} \and 
T. Abel\inst{10} \and 
K. Sadakane\inst{11}
}

\institute{Kapteyn Institute, University of Groningen,
Postbus 800, 9700AV Groningen, the Netherlands \and 
Institute of Astronomy, University of Cambridge,
Madingley Road, Cambridge CB3 0HA, UK \and 
Observatoire de Gen\`eve, EPFL CH-1290, Sauverny, Switzerland  \and 
Observatoire de Paris-Meudon, 92195 Meudon Cedex, France \and 
Department of Physics and Astronomy, University of Victoria, 3800 Finnerty Road, 
Victoria, BC, V8P 1A1, Canada \and 
University of Texas, McDonald Observatory, USA \and 
National Astronomical Observatory, 2-21-1 Osawa,
Mitaka, Tokyo 181-8588, Japan \and 
European Southern Observatory,
Karl-Schwarzschild str 2, D-85748 Garching bei M\"{u}nchen, Germany \and 
European Southern Observatory,
Alonso de Cordova 3107, Vitacura, Casilla 19001, Santiago, Chile \and 
Kavli Institute for Particle Astrophysics and Cosmology, Department of Physics and Stanford Linear Accelerator Center, Stanford University, 2575 Sand Hill Road, Menlo Park, CA 94044, USA  \and 
Astronomical Institute, Osaka Kyoiku University, Kashiwara-shi,
Osaka 581-8582, Japan.}

\date{Received / Accepted }

\abstract
{}
{As part of the DART project
we have used the ESO/2.2m Wide Field Imager  in conjunction with the
VLT/FLAMES\thanks{Based on FLAMES observations
collected at the European Southern Observatory,
proposal 171.B-0588} GIRAFFE spectrograph to study the detailed properties of  the resolved
stellar population of the Fornax dwarf spheroidal galaxy out to and
beyond its tidal radius. Fornax dSph has had a complicated evolution and contains significant 
numbers of young, intermediate age and old stars. We investigate the relation 
between these different components by studying their photometric, kinematic and abundance distributions.
}
{
We re-derived the structural parameters of the Fornax dwarf spheroidal  using
our wide field imaging covering the galaxy out to its tidal radius, and  analysed the spatial
distribution of the Fornax stars of different ages as selected from
Colour-Magnitude Diagram analysis.
We have obtained accurate velocities and metallicities
from spectra in the Ca~II triplet
wavelength region for 562 Red Giant Branch stars which have velocities consistent with
membership in Fornax dwarf spheroidal.}
{We have found evidence for the presence of at least three distinct
stellar components: a young population (few 100 Myr old)
concentrated in the centre of the galaxy, visible as a
Main Sequence in the Colour-Magnitude Diagram; an intermediate age
population (2-8 Gyr old); and an ancient population ($>$ 10Gyr),
which are distinguishable from each other kinematically, from the  metallicity
distribution and
in the spatial distribution of stars found in the Colour-Magnitude  Diagram.
}{
From our spectroscopic analysis we find that the
``metal rich'' stars ([Fe/H]$> -1.3$) show a less extended and more
concentrated spatial distribution, and display a colder kinematics
than the ``metal poor'' stars ([Fe/H$<-1.3$). There is tentative evidence
 that the ancient stellar population in the centre of Fornax
 does not exhibit equilibrium kinematics. This could be a sign of a  relatively recent
accretion of external material, such as the merger of another galaxy
or other means of gas accretion at some point in the fairly recent  past, consistent
with other recent evidence of substructure (Coleman et al. 2004, 2005).
}
{}

\titlerunning{DART Survey of Fornax dSph}
\authorrunning{G.Battaglia et al.}

\maketitle

\section{Introduction}
The Fornax Dwarf Spheroidal galaxy (dSph) is a relatively
distant companion of the Milky Way at high galactic latitude
({\it b}~=~$-$65.7$^o$).  It is located at a distance of 138 $\pm
8$~kpc \citep[e.g.][]{Mateo1998, Bersier2000}, with an heliocentric velocity 
v$_{\rm hel} =$ 53 $\pm 2$ km/s 
\citep{Mateo1991}, a luminosity $M_{\rm V} = -13.0 \pm 0.3$, and a 
central surface brightness $\Sigma_{\rm 0,V} = 14.4 \pm 0.3$ mag/arcmin$^2$ \citep{IH1995}.
The Fornax dSph is one of the most massive and luminous
of the dwarf galaxy satellites of our Galaxy, second only to
Sagittarius, with a total (dynamical) mass in the range $10^8$-$10^9$ \sm
 \citep{Mateo1991, Lokas2002, walker2006, Battaglia2006a}.
 
Similarly to Sagittarius, Fornax has its own globular cluster
population, but contrary to Sagittarius all of the Fornax globular
clusters appear to be metal poor \citep[e.g.][]{strader2003, letarte2006a}.
 Fornax is known to contain a large number of carbon
stars with a wide range of bolometric luminosities indicating
significant mass (and hence age) dispersion at intermediate ages 
\citep[e.g.][]{Demers1979, aaronson1980, stetson1998, azzopardi1999}. This is supported
by a well populated intermediate age sub-giant branch and red clump.
Fornax also contains a sizable population of RR Lyrae variable stars
with an average metallicity of [Fe/H]$_{\rm RR} \sim -1.6 \pm 0.2$
\citep{Bersier2002}. From detailed Colour-Magnitude
Diagram (CMD) analysis \citep[e.g.][]{stetson1998, buonanno1999, saviane2000, pont2004}
 it is clear that, unlike most dSphs in the Local Group, 
Fornax has had a long history of star formation, which has only
ceased few hundreds Myr ago. However, 
in common with most dSph galaxies, Fornax does not appear to have any HI gas at present 
\citep[e.g.][]{Young1999}. It is difficult to understand how Fornax lost all it gas. 
It is possible that, as consequence of the recent star 
formation, the gas is fully ionized, however detections of ionized gas 
are difficult.  

A recent proper motion estimate suggests that Fornax has a
low eccentricity polar orbit, that crossed the path of the Magellanic
Stream about 190~Myr ago \citep{dinescu2004}, and that the excess of
small scale structure in the Magellanic stream, tracing back along the
proposed orbit of Fornax comes from gas lost by Fornax during this
encounter.  The authors propose this encounter as a viable mechanism 
to make Fornax lose its gas and stop forming stars. However, previous 
proper motion measurements do not suggest such an encounter 
\citep{piatek2002}.

Fornax has been the centre of some attention recently for the
discovery, in photometric surveys, of stellar over-densities in and
around the system \citep{coleman2004, coleman2005, ol2006}.
These over-densities have been interpreted by Coleman et al. as {\it
shell structures} caused by the recent capture of a small galaxy by
Fornax. 

There have been previous spectroscopic studies of individual
stars in Fornax. 
Studies of kinematic properties
indicated a (moderate) central mass-to-light ratio $M/L \gtrsim 5
$($M/L_{\rm V}$)$_\odot$, possibly as high as 26 \citep[e.g.][]{Mateo1991}, 
and a global $M/L_{\rm V} \sim$ 10-40 larger than the $M/L_{\rm V}$ 
of the luminous component \citep[][]{walker2006}.
This is much higher than is typical for globular clusters, which have 
mass-to-light ratios in the range 1-3 \citep[e.g.][]{mac2005}, 
but
considerably lower than that found for many other dSphs such as Draco and
Ursa Minor \citep[$M/L >$ 100 ($M/L_{\rm V}$)$_{\odot}$, e.g.][]{kleyna2003, wilkinson2004}.  

There have
been two previous Ca~II triplet (CaT) studies of Red Giant Branch (RGB) stars in the
central region of Fornax (33 stars, Tolstoy et al. 2001; 117 stars, Pont et al. 2004),
 both using VLT/FORS1. There are also detailed
abundance studies of 3 field stars 
\citep{shetrone2003} and 9 globular cluster
stars \citep{letarte2006a} made with VLT/UVES. This  is about to be
extended dramatically by VLT/FLAMES \citep[$\sim 100$ stars,][]{letarte2006b}.
These spectroscopic studies conclude that the bulk of the stellar population is more
metal rich than could be inferred from the position of the Red Giant
Branch (RGB) in the CMD, with a peak at [Fe/H] $\sim -0.9$, a metal
poor tail extending out to [Fe/H] $= -2$ and a metal rich tail
extending beyond [Fe/H] $= -0.4$ with more than half the stars on the
RGB having ages $<$~4~Gyr. 

Here we present the first results of a study of the Fornax dSph from
the DART (Dwarf Abundances and Radial velocities Team) large programme
at ESO. 
The aim of DART is to analyse the chemical and kinematic behaviour of individual stars 
in a representative 
sample of dSphs in the Local Group, in order to derive their star formation and 
chemical enrichment histories, and explore the kinematic status of these objects 
and their mass distribution. Our targets are Sculptor (Tolstoy et al. 2004), 
Fornax and Sextans dSphs. From our large program at ESO we obtained for each of these 
galaxies: extended WFI imaging; intermediate resolution VLT/FLAMES spectra in the CaT region 
to derive metallicity and velocity measurements for a large sample of RGB stars covering an extented 
area; and VLT/FLAMES high resolution 
spectra in the central regions, which give detailed abundances over a large range of elements 
for $\sim 100$ RGB stars.  

In this work we present our results for both wide field imaging and 
intermediate resolution FLAMES spectroscopy for 562 RGB kinematic members of Fornax
out to the tidal radius. The relatively high
signal-to-noise (S/N $\approx$ 10-20 per \AA) of our data has enabled
us to derive both accurate metallicities ($\approx$
0.1 dex from internal errors) and radial velocities (to $\approx \pm$2
km/s). We can thus study the different stellar populations in
Fornax and how their metallicities and kinematic properties vary spatially 
with a dramatically larger sample than any previous study. Our study differs 
from Walker et al. (2006) in that we have both abundance and kinematics and 
this has been shown to be a crucial piece of information for disentangling 
surprisingly complicated galaxy properties (Tolstoy et al. 2004). 

In future works we will present our CaT calibration (Battaglia et al. 2006b), 
detailed kinematic modelling of Fornax dSph (Battaglia et al. 2006a) and 
accurate star formation history from our VLT/FLAMES high resolution abundances of Fornax 
(Letarte et al. 2006b).

\section{Observations and Data reduction}

\subsection{Photometry}
Our ESO Wide Field Imager (WFI) observations were collected between 2003-2005 
and initally included some earlier archival data from the central regions.
Table~\ref{tab:photo_obs} shows the journal of WFI observations taken in service mode early
in 2005 that were used to generate the final overall photometric and
astrometric catalogues.  Although the earlier WFI observations were used 
for target selection for some of the spectroscopic followup, all of the 
analysis and results presented in this paper are based on the 2005 service 
mode data.  This was taken in photometric conditions, with generally good
seeing ($<$1 arcsec), unlike most of the earlier photometry.
 
Image reduction and analysis was based on the pipeline processing software 
developed by the Cambridge Astronomical Survey Unit for dealing with imaging 
data from mosaic cameras.  Details of the pipeline processing can be found in 
\citet{Irwin1985}, \citet{irwin2001} and \citet{irwin2004}.  In broad outline,
each set of frames for a particular field is processed individually to remove 
gross instrumental artefacts (ie. bias-corrected, trimmed, flatfielded, and
defringed as necessary).  An object catalogue is then constructed and used to
derive accurate astrometric transformations to enable stacking of multiple 
images for each field in a common coordinate system.  The stacked images,
produced from a sequence of 3 dithered exposures, are then used to generate 
the final object catalogues, and the astrometric information is updated.

The astrometric solutions are based on a general Zenithal Polynomial projection
(e.g. Greisen \& Calabretta 2002), which for the ESO WFI is particularly simple
since the radial distortion is effectively zero.  Astrometric standard stars
were selected automatically from APM scans of UKST photographic survey plates
(http://www.ast.cam.ac.uk/casu) which in turn are calibrated with respect to 
TYCHO 2.  Each of the 8 detectors making up the mosaic is solved for 
independently using a 6 constant linear solution linking detector x,y to 
projected celestial coordinates.  Residuals in the fit per star are 
typically $\approx 200$mas, which is almost entirely due to the random error 
in the photographic astrometry.  Even in the outer parts of Fornax there are 
more than 100 plate-based standards per detector which means that any 
systematic errors in the solution are completely negligible ($<$100 mas).

Objects are morphologically classified on each stacked frame as either
noise artefacts, galaxies, or stars, primarily according to the properties
of the curve-of-growth light distribution using a series of aperture fluxes.  
At the same time aperture corrections for stellar profiles are derived at 
the detector level to enable accurate ($\approx$1\%) conversion to total 
fluxes.  

The internal gain correction, applied at the flatfielding stage, ensures
a common photometric system across the mosaic (to $\approx$1\%) and hence 
photometric calibration simplifies to finding a single magnitude zero-point 
for each science observation.  A correction for scattered light, which
affects the flatfielding in addition to the science images for WFI data, 
was also made.  This was directly derived from the standard star observations 
assuming a radially symmetric effect.  This is a well-known problem with the 
WFI \citep{eso2001} and the correction was made prior to the 
photometric calibration.

The first-pass photometric calibration is based on observations of 
photometric standard fields taken during the same nights as the target 
observations.  Since these observations were undertaken 
in service mode in mainly photometric conditions we used the default La Silla
extinction for these passbands to derive an airmass correction (the range 
in airmass and number of the standard star observations precluded any other 
option).  The colour equations to correct to the Johnson-Cousins system were 
taken from the ESO web pages.  Each photometric standard field observed 
contains between $\approx$10-50 suitable standards and provides a single 
zero-point with better than $\approx$1\% random error; assorted systematic 
errors (e.g. aperture corrections, flatfielding gradients, scattered light
correction) contribute at about the same level.

The zero-point trend over all the service observing is consistent within
the errors $\pm 1-2$\% with photometric conditions on all nights apart from 
the standard star observation for the night of 2005 Jan 31st.  Since no 
science target frames were taken during this night this data point was 
ignored and a constant zero-point adopted for both the V- and I-band
images.  (Note that this was also the night with the very poor seeing.)

As a final check on the overall calibration the overlap ($\approx 5$\%) 
between adjacent fields was used to check the internal consistency of the 
magnitude system and small $\approx \pm 2$\% adjustments were made to bring 
all the observations to a common scale.  This latter step ensures the whole 
system is on the same photometric zero-point, which is essential for analysis 
of large scale structure.  
The final sensitivity of all the images is the same within 0.1 magnitude.

The final step was to merge all the object catalogues in the V-
and I-band to produce a unique set of object detections for the whole area
studied.  In the case of multiple measurements for a given object the
one with the smallest error estimates were used. This catalogue forms the 
basis for all subsequent photometric analysis, and spectroscopic analysis.

\subsection{Spectroscopy}
We selected targets classified as stellar in our ESO/WFI photometry and 
with a position on the CMD consistent with an RGB star, but with a wide colour range to avoid 
biasing our sample in age or in metallicity.
We used VLT/FLAMES feeding the GIRAFFE spectrograph 
in Medusa mode, that allows the simultaneous allocation of   
132 fibres (including sky fibres) 
over a 25' diameter field of view \citep{Pasquini2002}. We used the GIRAFFE low resolution grating 
(LR8, resolving power R$\sim$ 6500), covering the wavelength range from 8206 \AA\,  to 9400 \AA, to  
obtain spectra for 7 different fields in Fornax dSph. 
These data have been reduced using the GIRAFFE pipeline 
(Geneva Observatory; Blecha et al. \citeyear{blecha2003}). 
The sky-subtraction and 
extraction of the velocities and equivalent widths of CaII triplet lines 
have been carried out using our own software 
developed by M.Irwin (for the details see Battaglia et al. \citeyear{Battaglia2006b}). 
Out to a total of 800 
targets, 69 have double measurements due to overlapping fields. 
We find that a S/N per \AA\ $>$ 10 is the minimum for accurate 
determination of velocity and equivalent width, thus
we exclude from our analysis the stars with S/N per \AA\ lower than 10. 
The typical error in velocity was found to be
 $\lesssim$ 2 \kms. We excluded from our 
analysis all the stars with an error in velocity larger than 5 \kms.
The r.m.s. velocities and equivalenth widths from the 35 stars with repeated measurements 
that passed our selection criteria 
are 2.2 \kms and 0.34 \AA\ respectively. The latter corresponds to an error in 
metallicity of 0.14 dex (see Sect.~4). For 27 out of 35 stars with 
double measurements we find agreement within the velocity errors. 
Only 2 out of 35 have velocities that differ by more than 2$\sigma$, 
however both are within 3$\sigma$.

The final sample was carefully checked to weed out any spurious objects 
(e.g. broken fibres, background galaxies, foreground stars, etc.). We removed 
6 objects because they were inadvertantly assigned to fibres which were not 
available during our observing runs; we found 
no background galaxies and removed 12 objects because the 
continuum shape or the presence of very broad absorption line was not consistent 
with what expected for RGB stars spectra. Excluding the objects that did not meet our  
S/N and velocity error criteria, our final sample of acceptable measurements consists of 641 stars. 

\section{Results: Photometry}

\begin{figure}
\begin{center}
\includegraphics[width=140mm, angle=0]{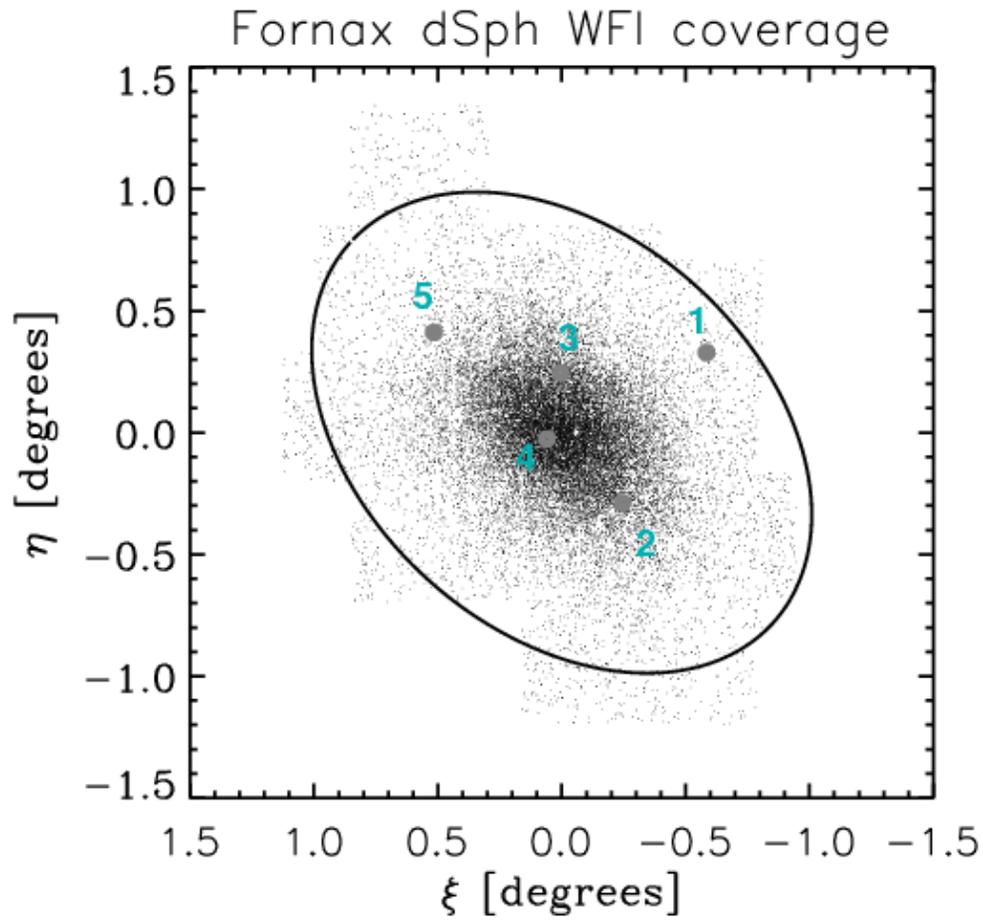}
\caption{ESO/WFI coverage for Fornax dSph. The inner shell-like feature detected by Coleman et al. (2004) 
is located at $\xi \sim$0.12, $\eta \sim -$0.20 and elliptical radius of 0.33 deg. The circles indicate 
the location of Fornax globular clusters and the ellipse shows Fornax tidal radius (ellipse parameters 
derived in this work, see Table~\ref{tab:par}). The North is at the top, the East on the left-hand side. 
The physical scale at the distance of the Fornax dSph is 1 deg $=$ 2.4 kpc.}
\label{fig:coverage}
\end{center}
\end{figure}

In this section we present the results of our wide field, 
ESO/WFI photometry of Fornax, out to  
its nominal tidal radius (see Fig.\ref{fig:coverage}), and going down to $V=$ 23 and 
$I=$ 22 (M$_{\rm I}=$ 1.2).

\subsection{The structure of Fornax}

Previous studies have revealed the spatial structure 
of Fornax to be far from regular. Hodge (1961), using photographic plates,
 found that the ellipticity (defined as $e= 1- b/a$, where $b$ and $a$ 
are respectively the semi-minor and major axes of the galaxy) 
increases with radius, from a value of 0.21 in the central regions to 0.36 
in the outer parts. 
\citet{IH1995} noticed that in the inner regions the isopleths 
showed departures from elliptical symmetry and that they were more closely spaced 
on the east side of the major axis than in the west, as already noticed by Hodge (1961). 
The asymmetry was found to be centred close to cluster 4 ($\xi \sim 0.05$, $\eta \sim -0.1$ in 
Fig.~1). The ellipticity was again found to 
increase with radius.
\citet{coleman2005}, using only RGB stars, found that the position 
of the centre shifts 3$'$ towards west within a radius of 20$'$ and 4$'$ towards south 
at all radii. For radii larger than 35$'$ they find an opposite trend for the ellipticity, 
i.e. the outer contours become more circular. 

\begin{figure}
\begin{center}
\includegraphics[width=60mm]{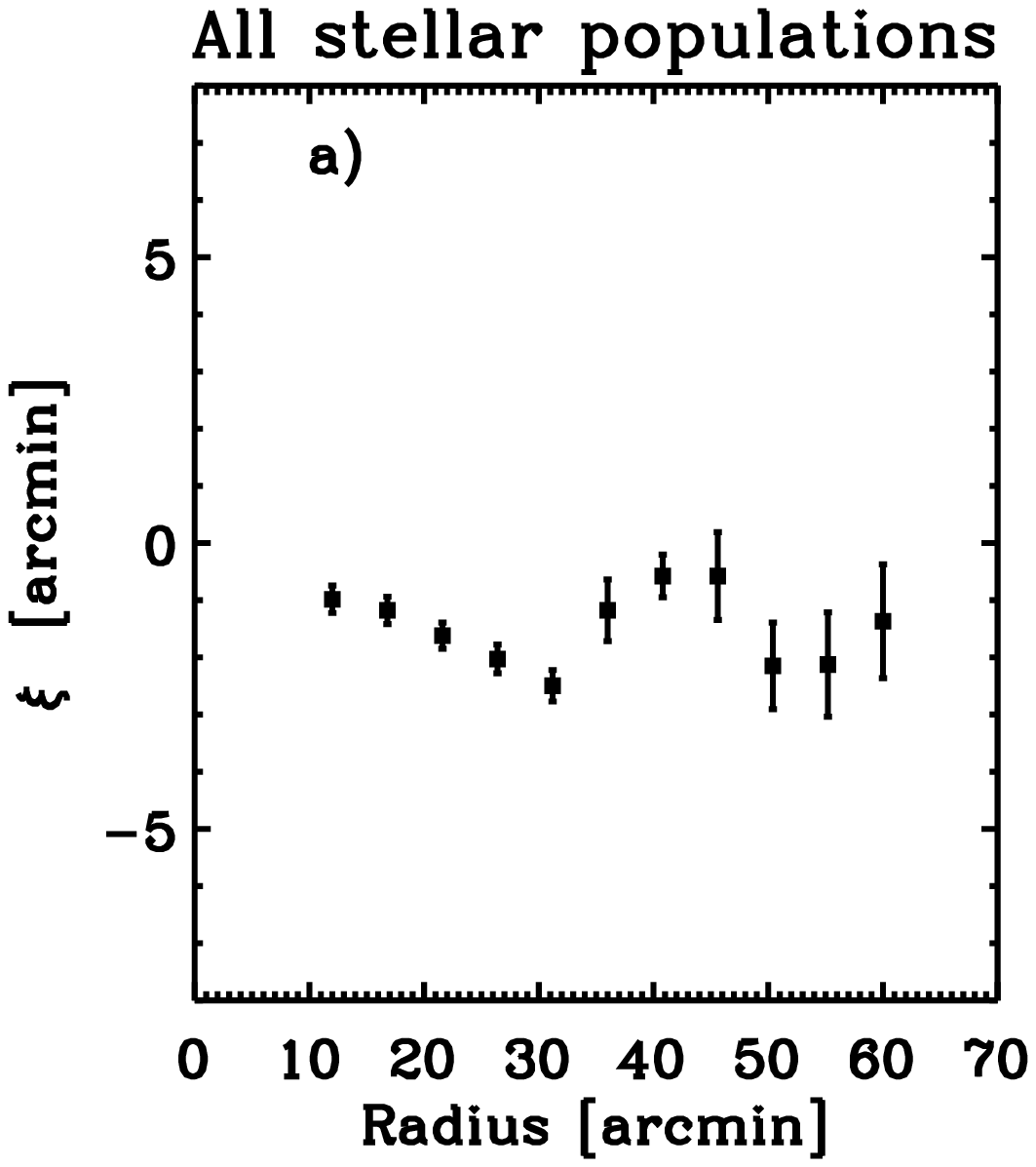}
\includegraphics[width=60mm]{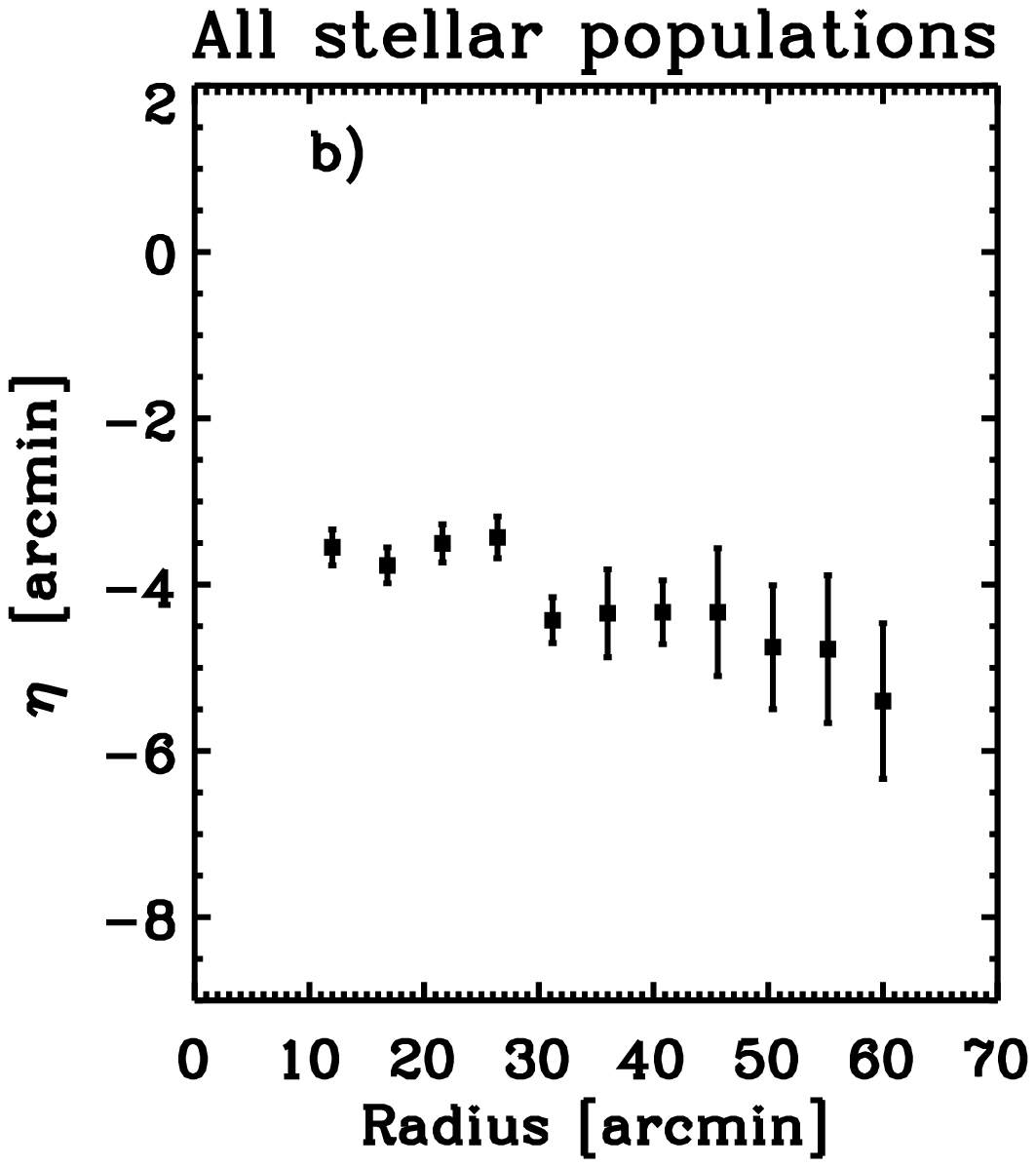}
\includegraphics[width=60mm]{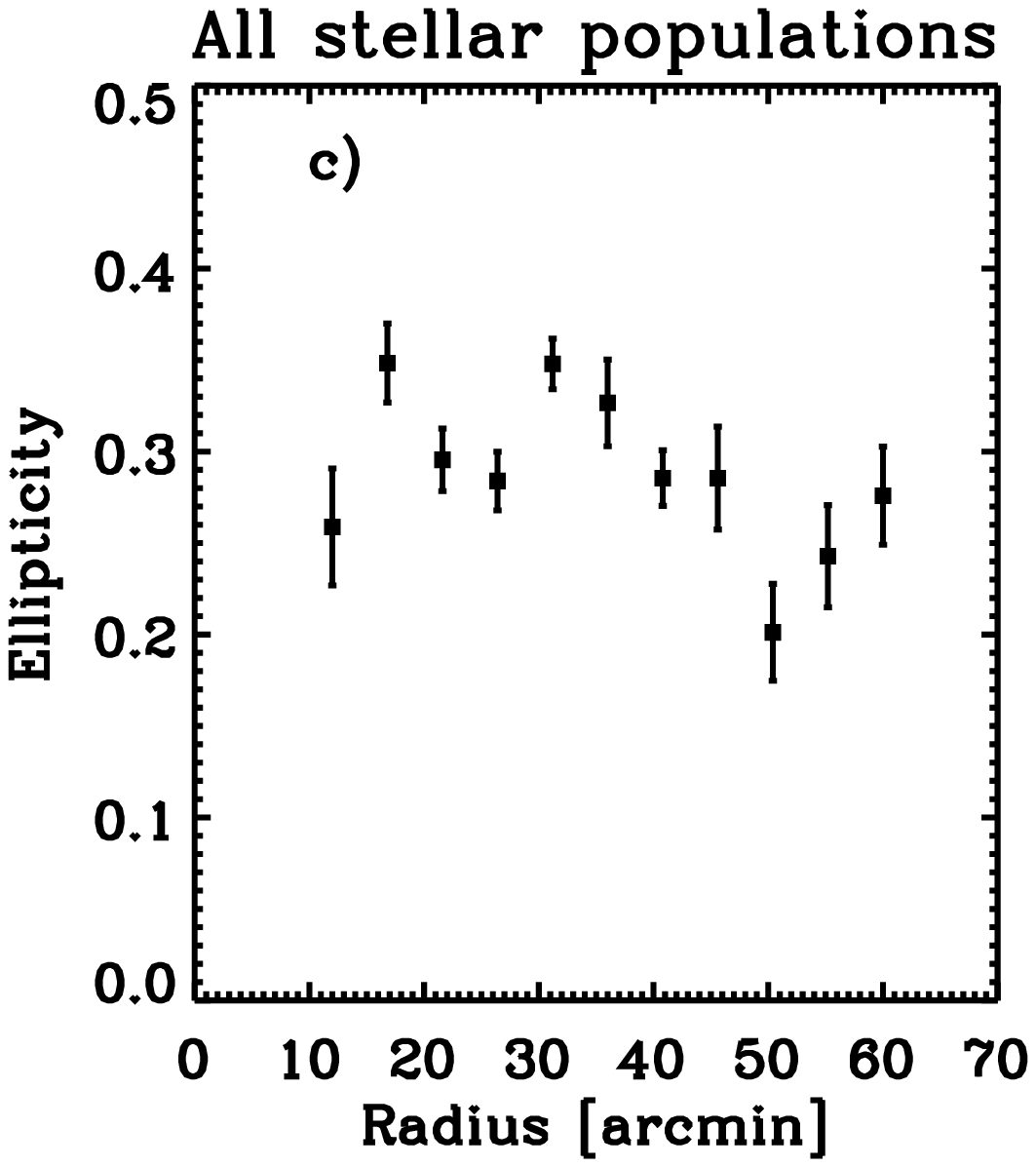}
\includegraphics[width=60mm]{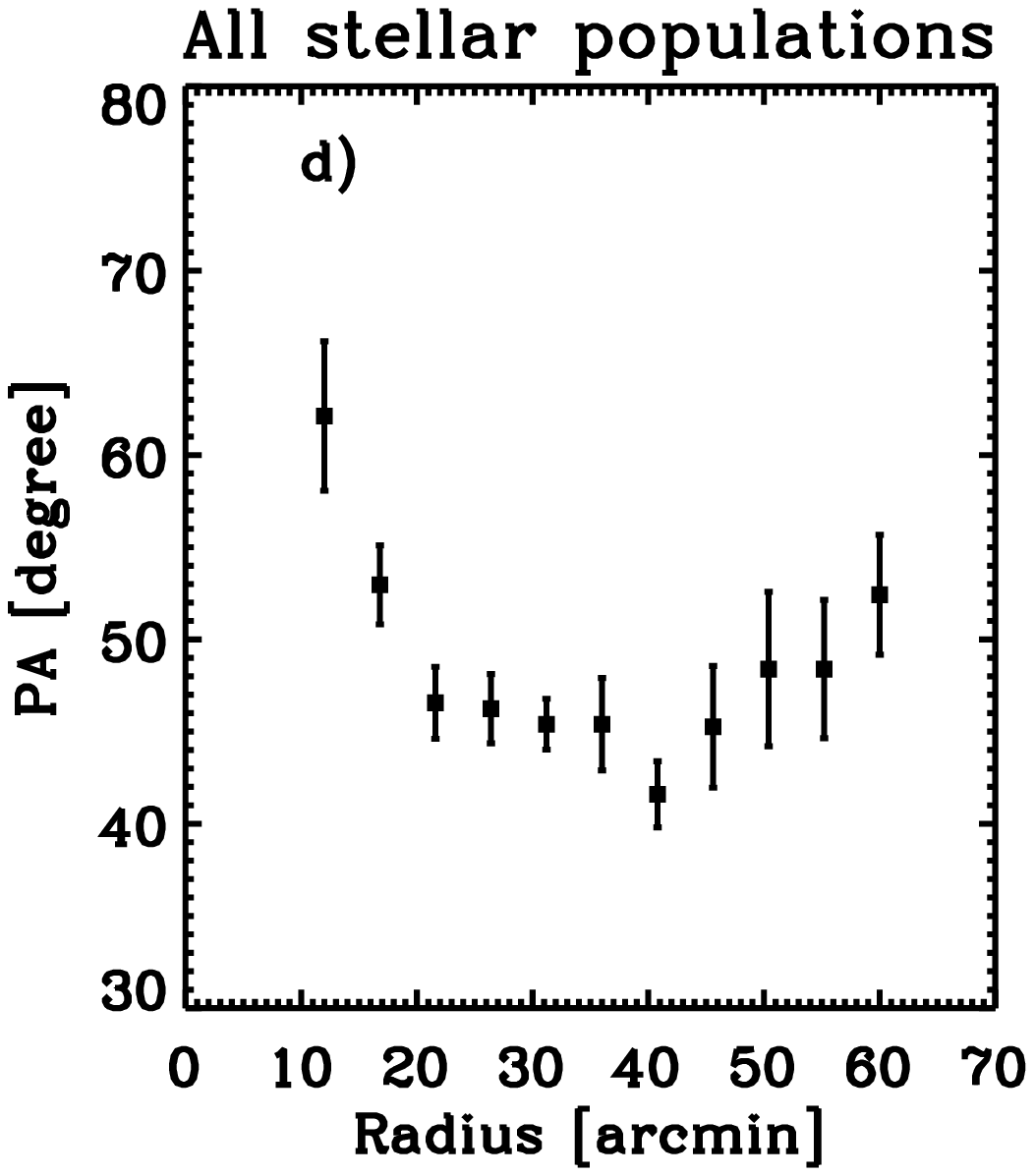}
\caption{The variation with radius of the central position ($x$ and $y$ coordinate, a) and b) respectively), c) ellipticity and d) position angle for the entire stellar population of Fornax, from ESO/WFI imaging data. The variation of 
the central position is with respect to the value listed in Mateo (1998). We find that the centre is at 
$\alpha =2^h 39^m 52^s$ and $\delta= -34^{\circ} 30' 49''$, 
west ($-1.5' \pm 0.3'$) and south ($-3.8' \pm 0.2'$) with respect to the values listed in Mateo (1998).}
\label{fig:centre_all}
\end{center}
\end{figure}

Due to the variety of previous estimates, sometimes differing significantly from each other, we 
decided to re-derive the structural parameters of Fornax dSph. We divided the 2D 
spatial distribution of the stellar objects in our photometry 
(Fig.~1) into a grid with a ``pixel'' 
width of 0.02 deg and associated to each pixel the number of stars it contained. 
We then used the IRAF task ELLIPSE to derive the variation with radius of the  
central position, ellipticity and position angle (Fig.~\ref{fig:centre_all}), 
by fitting ellipses at radii between 15$'$ and 
60$'$ with a $\sim5'$ spacing.
We find that the centre is at ($\alpha_{\rm J2000}$, $\delta_{\rm J2000}$) 
$=$ ($2^h 39^m 52^s$, $-34^{\circ} 30' 49''$), 
west ($-1.5' \pm 0.3'$) and 
south ($-3.8' \pm 0.2'$) with respect to the values listed in Mateo (1998).  
The centre now appears to be located very close to cluster 4. 
With this new analysis
 the ellipticity, average value 0.30$\pm$0.01, does not show any trend with radius.   
The position angle (P.A.) appears to be larger in the inner 15$'$ and then to become 
approximately constant with radius. A larger position angle for the inner region could be 
caused, for example, by the presence of cluster 4 or, more likely, by 
the presence of young stars 
with a markedly different distribution (see Sect.~3.3). In our subsequent analysis we adopt 
the average value of the P.A. to be 46.8$^{\circ}\pm1.6^{\circ}$, consistent 
with previous estimates. Our re-derived structural parameters are listed in Table~\ref{tab:par}. 
With the exception of the position of the centre of the galaxy, 
all the derived parameters are consistent with the values listed by Mateo (1998).

\begin{figure}
\begin{center}
\includegraphics[width=100mm]{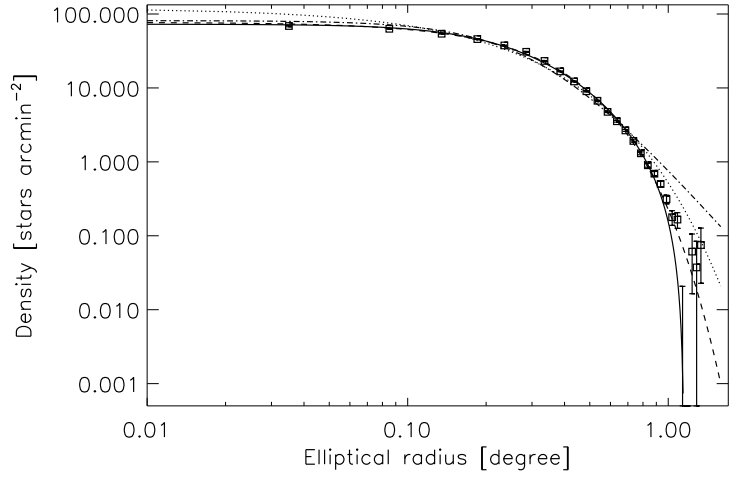}
\caption{The surface density profile for Fornax dSph galaxy with overlaid 
best-fitting King, Sersic, exponential and Plummer models 
(solid, dashed, dotted, dash-dot lines, respectively). The Galactic stellar contamination, 
0.78$\pm$0.02 stars arcmin$^{-2}$ calculated from a weighted average of the points beyond 1.13 deg (68$'$), has been 
subtracted from each point. The error bars are obtained by summing in quadrature the errors from Poisson statistics and 
the error in Galactic stellar contamination. Our best fit to the data is a Sersic profile with Sersic radius, 
$R_{\rm S}= 17.3 \pm 0.2$ arcmin and $m=$ 0.71. The best-fitting parameters for the different models are 
summarized in Table~\ref{tab:par_oldint}.}
\label{fig:density}
\end{center}
\end{figure}

We also rederived the 
surface density profile for Fornax in bins 
of 3$'$ out to a radius of 83$'$ (see Fig.~\ref{fig:density}).
The contamination by
field star density, from a weighted average of the 
points beyond 68$'$, is 0.78$\pm$0.02 stars arcmin$^{-2}$. 
From previous estimates of the tidal radius \citep[e.g.][]{IH1995} 
we know that it is unlikely that our data extend out to  
a pure background region thus our determination of the Galactic contamination 
is of necessity an overestimate due to our limited spatial coverage.
We compared this Galactic contamination subtracted profile   
to several surface brightness models using a least-square fit to the data. 
We used an empirical King profile \citep{king1962}, 
an exponential profile, 
a Sersic profile \citep{sersic1968} and a Plummer model \citep{plummer1911}. 

The King model has been extensively used to describe the 
surface density profile of dSphs.
\begin{equation}
I_{\rm K}(R)= I_{\rm 0,K} \ \left( \frac{1}{\sqrt{1 + \left(\frac{R}{r_c}\right)^2}} - 
\frac{1}{\sqrt{1 + \left(\frac{r_t}{r_c}\right)^2}} \right)^2
\end{equation}
It is defined by 3 
parameters: a characteristic surface density, $I_{\rm 0,K}$, core radius, $r_{\rm c}$ 
and tidal radius, $r_{\rm t}$. The last parameter is defined 
as the distance, along the line connecting the galaxy centre and 
the centre of the host galaxy, where at the perigalacticon 
passage a star is pulled neither in nor out. This radius 
is set by the tidal field of the host galaxy. 
Thus, an excess of stars in the outer parts of dSphs with respect 
to King models have been interpreted as tidally stripped stars, assuming that the King 
profile is the correct representation of the data. Excess stars beyond the 
King tidal radius have been found in Fornax \citep[e.g.][]{coleman2005} and in a number of other Local Group 
dSphs, e.g. Carina \citep{majewski2005}, Draco \citep{wilkinson2004, munoz2005}. 

Another possibility to explain the excess of stars in the outer parts 
of dSphs is that the King model 
does not provide the best representation of the data at large radii. 
The Sersic profile is known to provide a good empirical formula to fit the projected light distribution of 
elliptical galaxies and 
the bulges of spiral galaxies \citep[e.g.][]{caon1993, caldwell1999, graham2003, trujillo2004}: 
\begin{equation}
I_{\rm S}(R)= I_{\rm 0,S} \ exp \left[ - \left(\frac{R}{R_{\rm S}}\right)^{1/m}\right]
\end{equation}
where $I_{\rm 0,S}$ is the central surface density, $R_{\rm S}$ is a scale radius and $m$ is the 
surface density profile shape parameter. 
The de Vaucouleurs law and exponential are recovered with $m=4$ and $m=1$, respectively; 
 for $m < 1$ the profile is steeper than an exponential; $m=1/2$ gives a Gaussian. The light 
profile of some dSphs in the Local Group are found to be best-fitted with 
$m < 1$ \citep[e.g.][]{caldwell1999}.  
Physical justifications for this law have been proposed \citep[see for instance][and references therein]{graham2003}, 
however none is widely accepted. 

In our analysis King and Sersic profiles reproduce the data 
much better than a Plummer or an exponential profile. Our best fit 
is a Sersic profile with Sersic radius, $R_{\rm S}=$ 17.3$\pm$ 0.2 
arcmin, and $m=$ 0.71, thus steeper than an exponential (Fig.~3).  

For the best-fitting King model we find a core radius  
$r_{\rm c}=$ 17.5$\pm$0.2 arcmin and a tidal radius $r_{\rm t}=$ 69.4$\pm$0.4 arcmin. 
The core radius determined here is 4$'$ larger than previous estimates 
whilst the tidal radius is 
in agreement with the value of 71$\pm$4 arcmin from IH95. The slightly
smaller value 
we obtain for the tidal radius is likely due to our overestimated field star density.

Figure~\ref{fig:density} shows that in our determination of the density profile 
there is an excess of stars with respect to the King profile (solid line) for radii 
larger than 1 deg. At this distance the Sersic profile (dashed) best follows the data. 
Both exponential and Plummer profiles (dotted and dash-dot) 
over-predict the surface density at $r<0.1$ deg and 
at radii larger than $\sim$ 0.8 deg, and under-predict it at intermediate distances. 
A summary of the best-fitting parameters is given in Table~\ref{tab:par_oldint}.

\subsection{The Colour-Magnitude Diagram}

It is well estabilished that the Fornax dSph has had a very complex star formation 
history \citep[e.g.][]{stetson1998, buonanno1999, saviane2000}. 
Stetson et al. (1998), in a photometric study covering a field of $~$1/3 deg$^2$ 
near the centre of Fornax, found that the majority of stars fall into 
three distinct stellar populations: an old one represented by HB and RR Lyrae stars; 
an intermediate age population of He-core-burning clump stars (Red Clump, RC); 
and a young population, indicated by the presence of main sequence stars. 
Fornax displays a broad 
RGB, classically interpreted as due to a large spread in metallicity.

\begin{figure}
\begin{center}
\includegraphics[width=100mm]{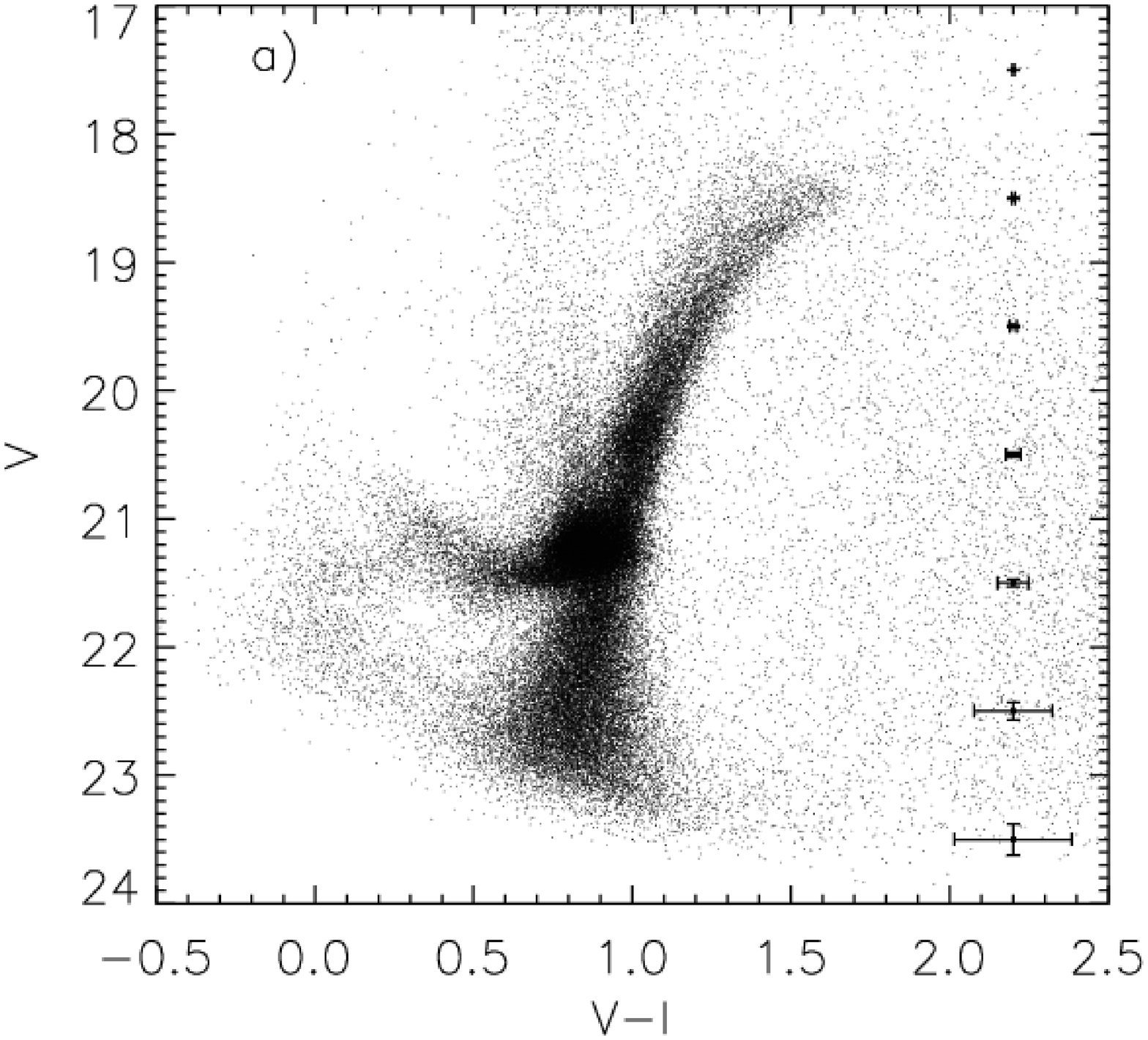}
\includegraphics[width=100mm]{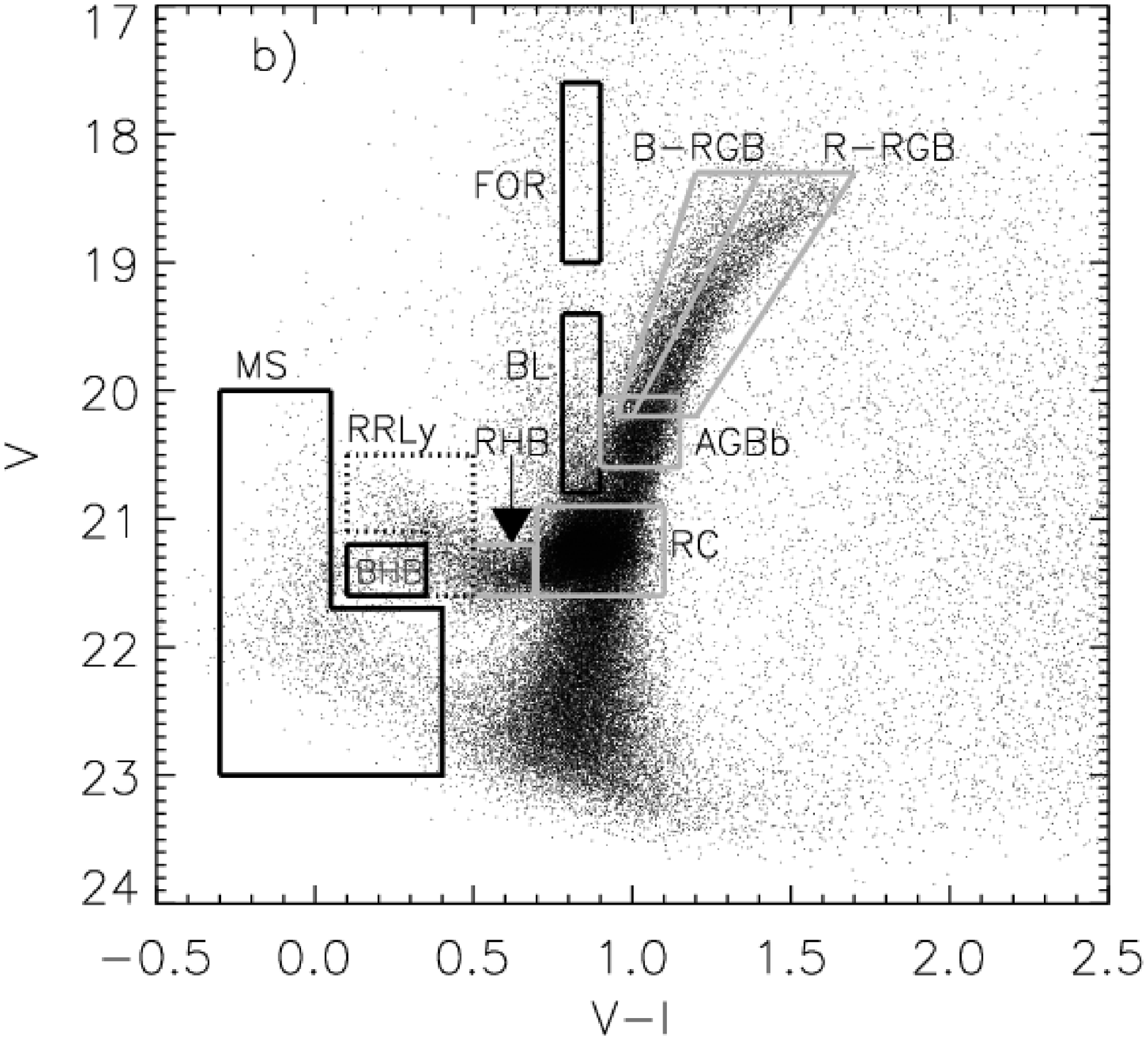}
\caption{CMD for objects classified as stellar 
from our ESO/WFI imaging covering the whole extent of the Fornax dSph. 
The boxes in panel b) indicate the regions 
we used to select different stellar populations (MS: main sequence; BL: Blue~Loop; 
FOR: foreground; BHB: Blue Horizontal Branch; RRLy: RRLyrae; RHB: Red Horizontal Branch; 
B-RGB: Blue-Red Giant Branch; R-RGB: Red-Red Giant Branch; AGBb: AGB bump; RC: Red Clump).}
\label{fig:cmd}
\end{center}
\end{figure}

\begin{figure}
\begin{center}
\includegraphics[width=80mm]{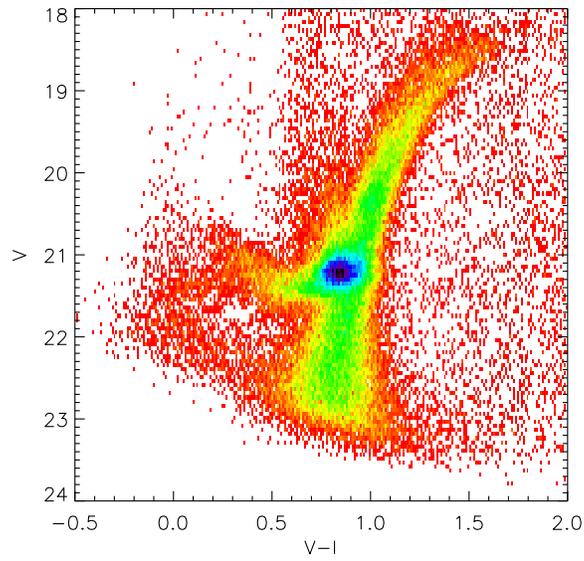}
\caption{Here we show the Hess diagram for Fornax dSph CMD from ESO/WFI imaging, where 
the structures, and their relative intensity, in the different regions of the CMD are visible. 
Each cell is $\Delta (V-I) \times \Delta V = 0.01 \times 0.05$. 
A darker tone represents an higher density of 
stars; zero-density is set to white. This allows us to see the structure in the Red Clump and 
Red Horizontal Branch.}
\label{fig:hess}
\end{center}
\end{figure}

Figure~\ref{fig:cmd}a shows our WFI 
Colour-Magnitude diagram for Fornax within the tidal radius 
($\sim 2^{\circ} \times 2^{\circ}$);
it contains 73700 stars, including a quantity of foreground stars, which 
lie in a fairly constant density sheet redwards of $V-I =$ 0.6.  
We notice the following features (highlighted in Fig.~\ref{fig:cmd}b, and 
directly visible in the Hess diagram in Fig.~\ref{fig:hess}):

\subsubsection{Ancient stars ($>$ 10 Gyr):} 
These stars are to be found in:
\begin{itemize}
\item A red horizontal branch (RHB), extending from $V\sim$21.2 to  $V\sim$21.6
and from $V-I \sim$0.5 to  $V-I \sim$0.7. 
Hints of a blue horizontal 
branch (BHB) from $V \sim$21.2 to  $V\sim$21.6 and from $V-I \sim$0.05 to $V-I \sim$0.35.
Between the BHB and RHB we can see the instability strip where RR Lyrae 
variable stars are to be found.
\item A broad Red Giant Branch (RGB) with distinct blue (B-RGB) 
and red (R-RGB) components. The RGB branch contain stars with ages $>$ 10 Gyr, as well as intermediate 
age stars (2-8 Gyr), as will be discussed in Sect.4.2. 
There is a  hint of bifurcation below the tip of the RGB.
\end{itemize}

\subsubsection{Intermediate age stars (2-8 Gyr):}
These stars are to be found in:
\begin{itemize}
\item An ``AGB bump'', from $V \sim$ 20.05 to $V \sim$ 20.6 and 
from $V-I \sim$0.9 to $V-I \sim$1.15. 
A similar feature was detected by \citet{gallart1998} in the LMC and 
called an ``AGB bump'', and assumed to occur at the beginning of 
the AGB phase.
\item A red clump (RC), that hosts the bulk of detected stars ($\sim 30$\%) in our WFI CMD.  
It extends from $V \sim$ 20.9 to $V \sim$ 21.6 and 
from $V-I \sim$0.7 to $V-I \sim$1.1. From theoretical modelling of RC stars, 
\citet{caputo1995} found that 
it is possible to 
estimate the age of the dominant stellar component from the extension 
in luminosity of the RC. They found that the larger the spread in luminosity the younger is  
the RC population. The predictions in \citet{caputo1995}
are made using models of metallicity Z=0.0001 and 0.0004, 
which is lower than the bulk of Fornax population (Z$\sim$ 0.002). 
Assuming a distance to Fornax of 138 kpc, an extinction in $V$ band of 
0.062 \citep{schlegel1998} and a colour excess E($V-I$)$\simeq$0.026
the upper and lower $M_{\rm V}$ of the RC are 0.14 and 0.84. Using the 
above mentioned models, this translates into 
an age of 3-4 Gyr for the dominant stellar population, assuming no strong variation due to metallicity. 
We also derive the age of the dominant RC stellar population by overlaying to the observed CMD 
Padua isochrones (Girardi et al. 2000), which have a mass grid fine enough to resolve the red clump structure.  
Even though isochrones predictions for colour and magnitude of the red clump feature are valid for single stellar populations, 
they will provide a reasonably accurate estimate 
of the mean age of a composite stellar population such as Fornax. We find that the observed mean magnitude 
of the RC feature ($V=$ 21.2, see Fig.~\ref{fig:hess}) 
is reproduced assuming a dominant stellar population of 2-4 Gyr old and metallicity 
Z in the range 0.001-0.004. 
\end{itemize}

\subsubsection{Young stars ($<$ 1 Gyr):}
These stars are found in:
\begin{itemize}
\item A plume of young Blue~Loop (BL) stars from $V \sim$ 19.4 to $V \sim$ 20.8 and 
from $V-I \sim$0.78 to $V-I \sim$ 0.9. Blue~Loop stars are produced during 
the He-core burning phase for stars with mass $\gtrsim$ 2-3 \sm and ages from few million years to 1 Gyr. 
The position (colour) of this feature is strongly dependent on metallicity 
\citep[e.g.][]{girardi2000} . 
\item A main sequence extending from $V\sim$ 20 to $V\sim$ 23 at \vi$\sim 0$, 
showing that very recent 
star formation (less than 1 Gyr ago) has occurred in Fornax. 
The maximum luminosity of this feature indicates that there 
are stars as young as 200-300 Myr in Fornax.  
\end{itemize}

\subsection{The spatial distribution of different stellar populations} 

The spatial distribution of different stellar components in a galaxy
gives an indication of how and where 
the star formation took place at different times and how long the gas was retained
as a function of radius. 

Previous studies of dSphs 
in the Local Group \citep[e.g.][]{hurley-keller1999, majewski1999, harbeck2001, 
tolstoy2004} have shown that 
HB properties vary as a function of the distance from the centre 
in several dSphs: in general 
the RHB is more centrally concentrated than 
the BHB. This indicates the dominance of old and/or metal poor 
stars in the outer regions and younger and/or more metal rich stars in the inner regions.

Fornax contains both old (e.g. HB), 
intermediate (e.g. RC), and young stellar components (MS and Blue~Loop 
stars). Previous photometric studies of individual stars, which have been 
limited to the central regions of Fornax dSph, 
have found indications that the intermediate age stars have a more concentrated 
distribution than the ancient stars \citep[]{stetson1998, saviane2000}. 
Stetson et al. (1998) found that the MS stars displayed a centrally concentrated asymmetric distribution, 
with a major axis tilted 
$\sim$ 30$^{\circ}$ with respect to the main body of Fornax.

\begin{figure}
\begin{center}
\includegraphics[width=0.5\textwidth]{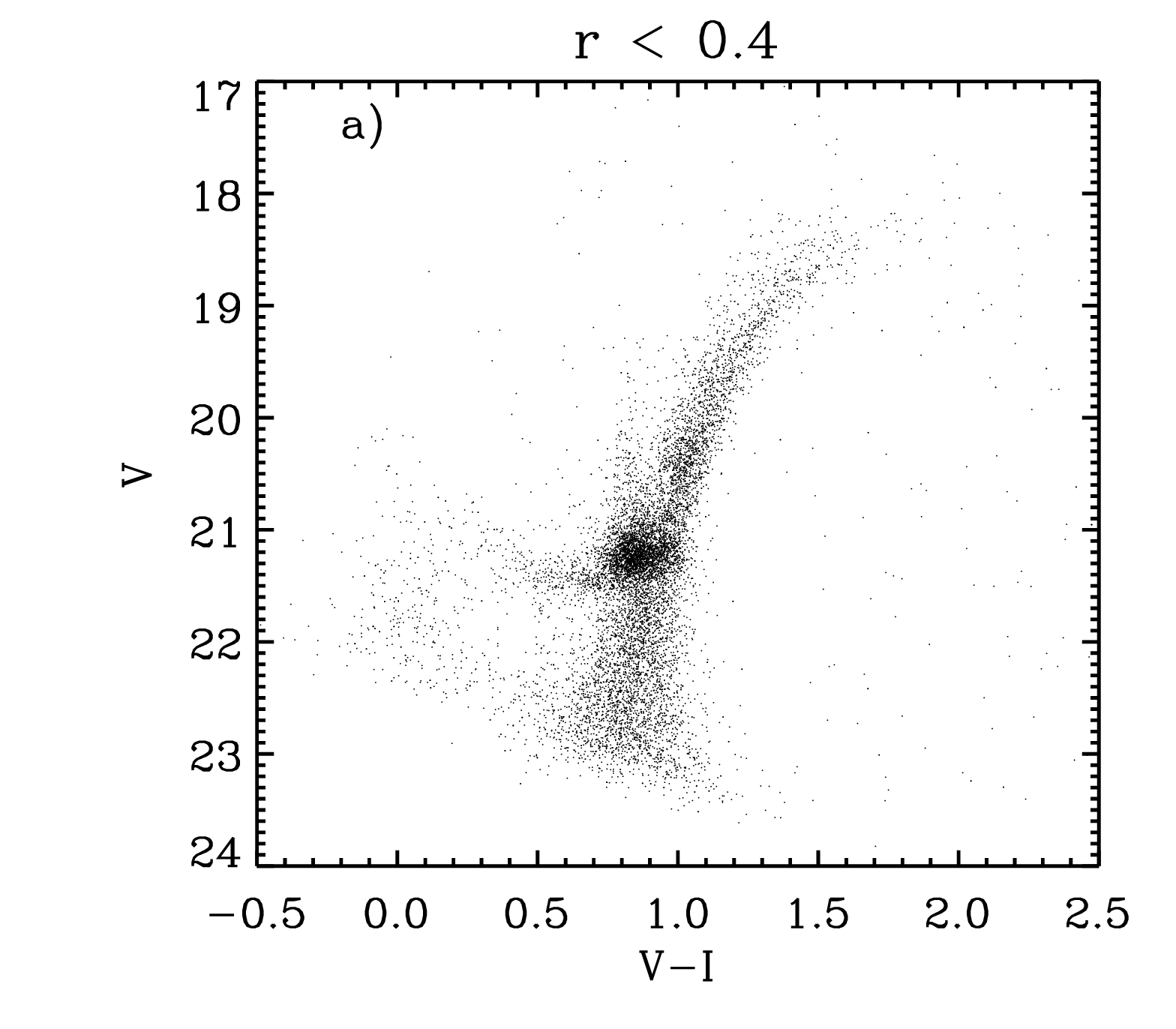}
\includegraphics[width=0.5\textwidth]{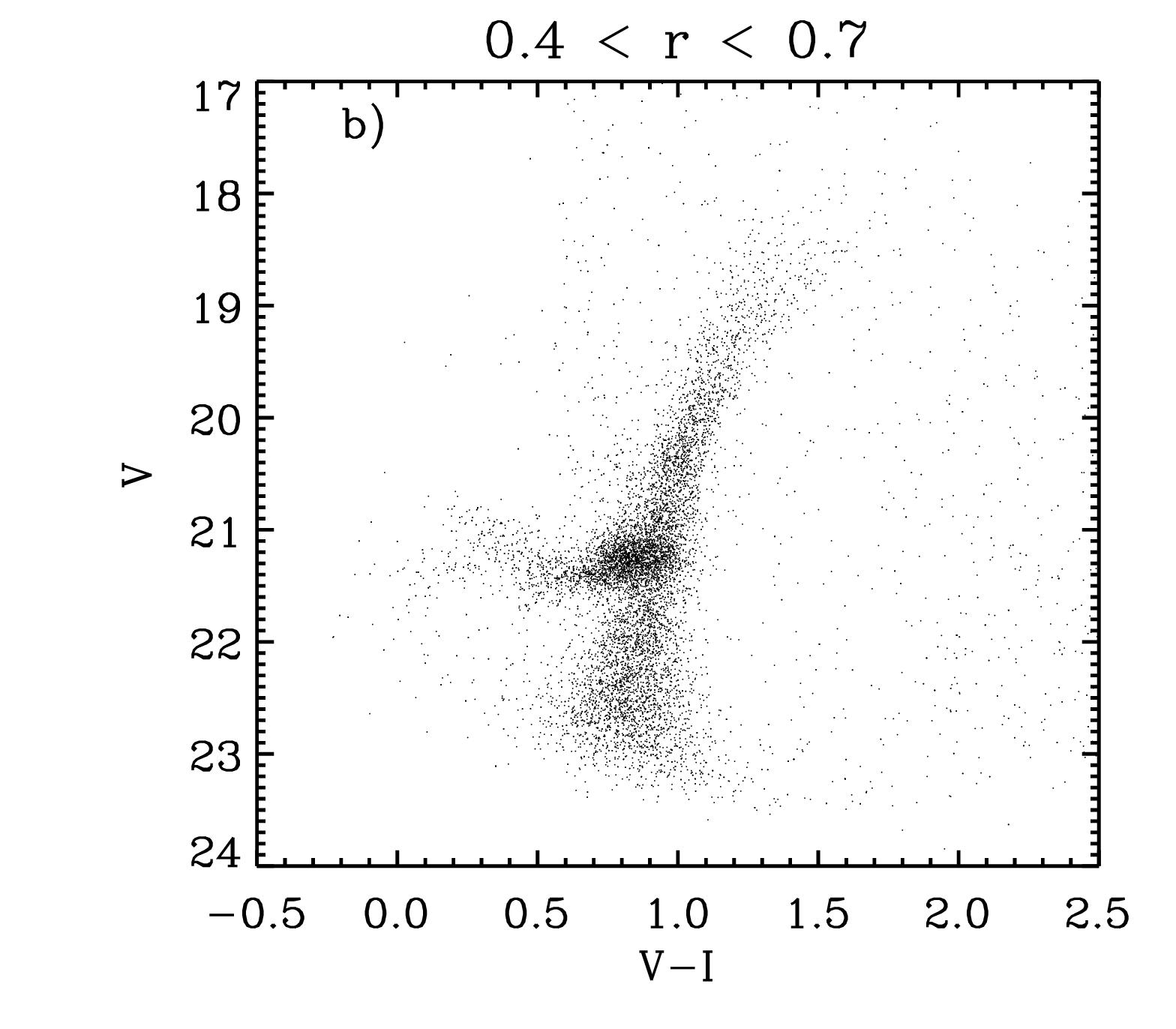}
\includegraphics[width=0.5\textwidth]{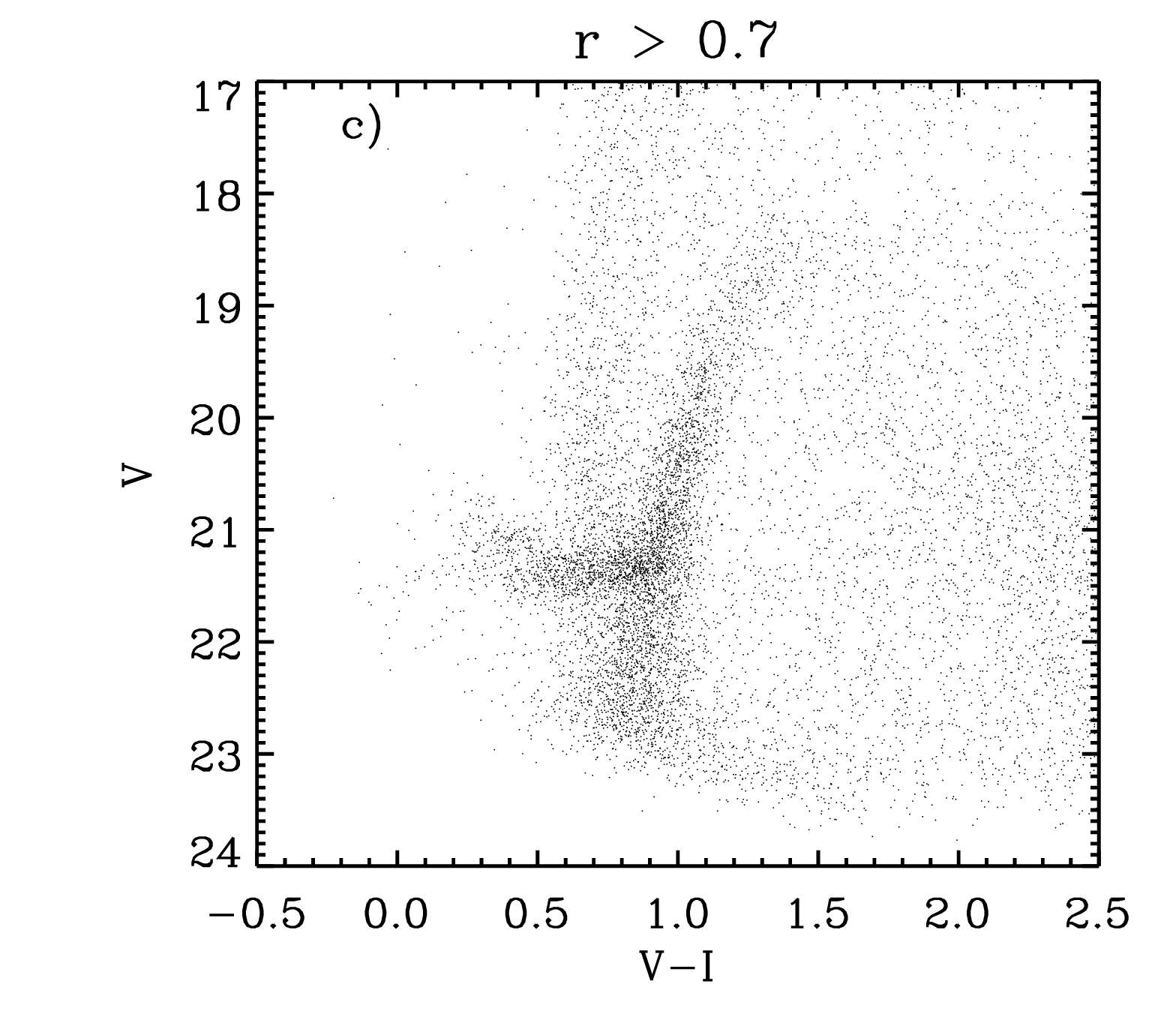}
\caption{ESO/WFI CMDs for the Fornax dSph at radii $r<$ 0.4 (a), 0.4 $<r<$ 0.7 (b), $r>$ 0.7 deg (c). 
We have normalized the number of stars in the 3 panels to allow a more meaningful comparison. 
Note the change in stellar population as function of radius: 
the young population (main sequence and Blue~Loop stars) is absent beyond $r>0.4$ deg, whilst 
the ancient population becomes more dominant, as can be seen for example from the more visible BHB at $r>0.4$ deg, the bluer average colour of the RGB, and the less extended RC in $V$ magnitude. 
}
\label{fig:cmd_r}
\end{center}
\end{figure}

Our WFI data, covering the entire extent of Fornax, allow us to analyse the 
spatial distribution of the resolved stellar population of Fornax out to its nominal tidal radius.  
We plot a CMD for 3 different spatial 
samples, i.e. for stars found at elliptical radius $r<$ 0.4 deg, 0.4 deg $< r <$ 0.7 deg, 
and $r>$ 0.7 deg (Fig.\ref{fig:cmd_r}). Clear differences are visible between the inner and the outer 
regions: the main sequence and Blue~Loop stars (young) are absent 
in the outer regions. 
The extent in $M_{\rm V}$ of the RC decreases with the distance 
to Fornax centre; 
this means that the age of the dominant stellar component in this feature 
is older in the outer parts \citep{caputo1995}. 
We also notice that the BHB, which is almost hidden by the main 
sequence in the inner region, becomes more clearly visible the further out we go, 
because the MS disappears. Finally, 
we can see that the RGB changes: the red side, generally younger and more metal rich than 
the blue side, becomes less evident in the outer regions. 

Below we analyse 
the spatial distribution of the different components, through their location 
in the field of view and the trend of the cumulative fraction with radius.

\subsubsection{Ancient populations}

\begin{figure}
\begin{center}
\includegraphics[width=0.5\textwidth]{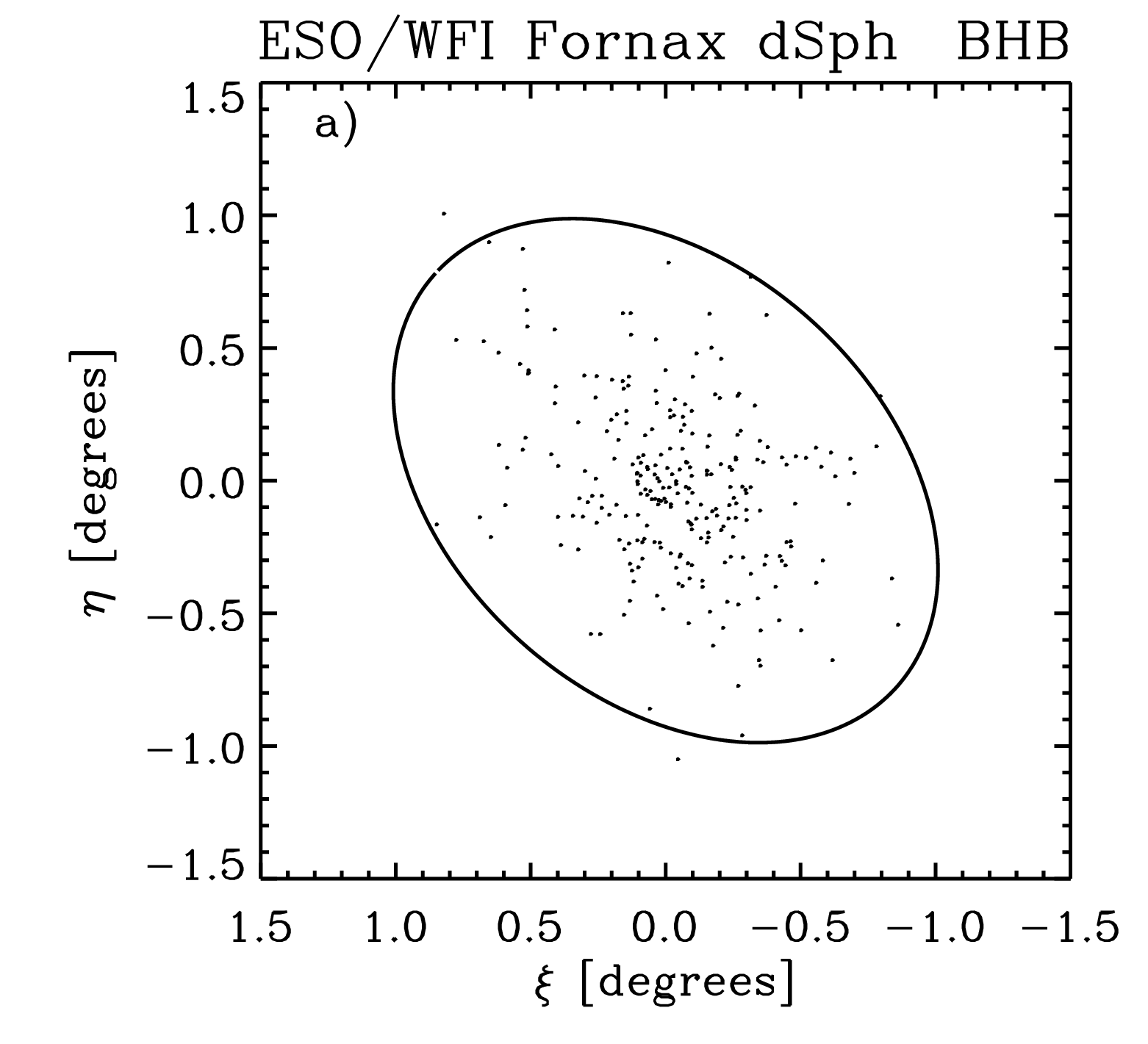}
\includegraphics[width=0.5\textwidth]{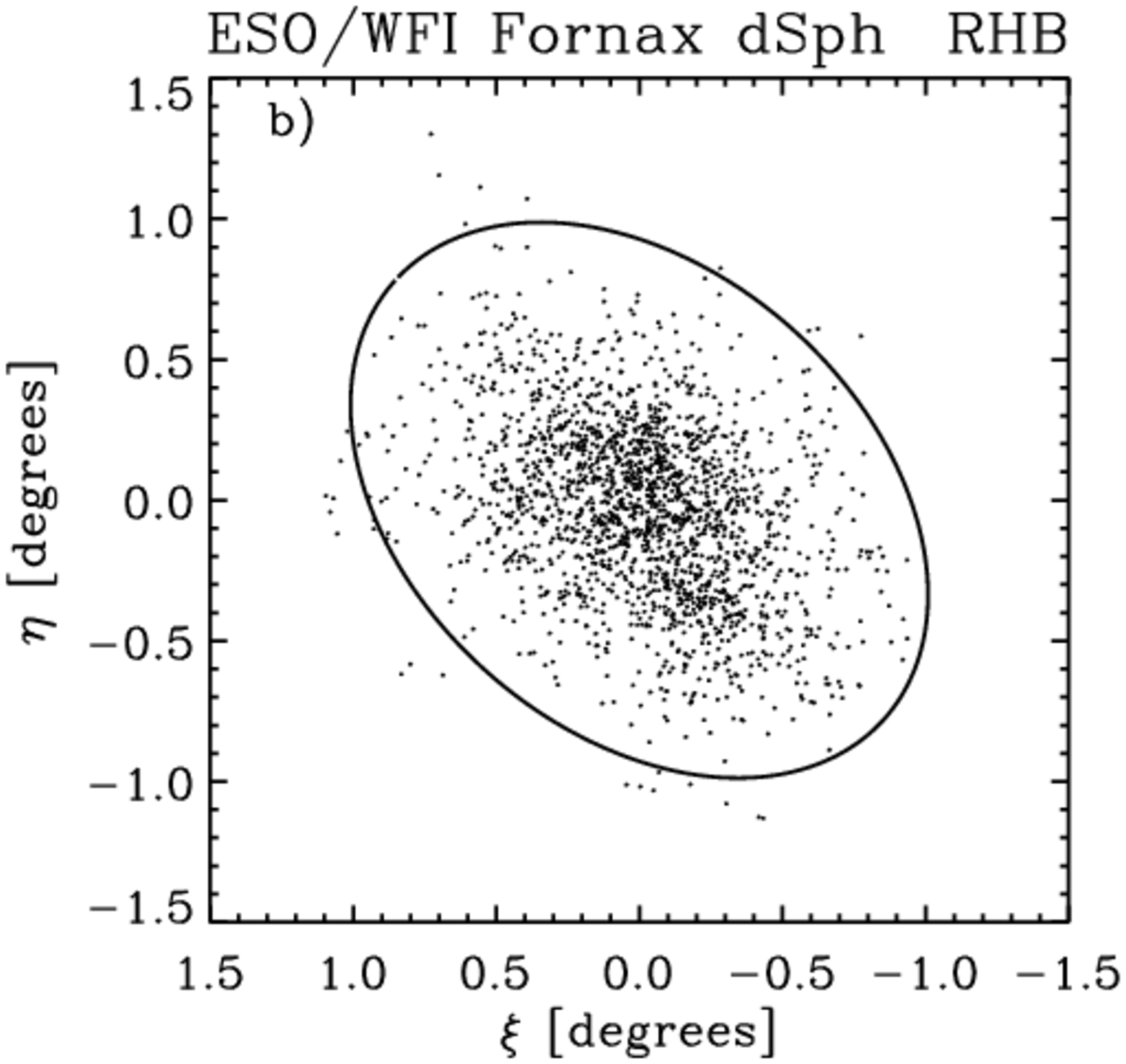}
\includegraphics[width=0.5\textwidth]{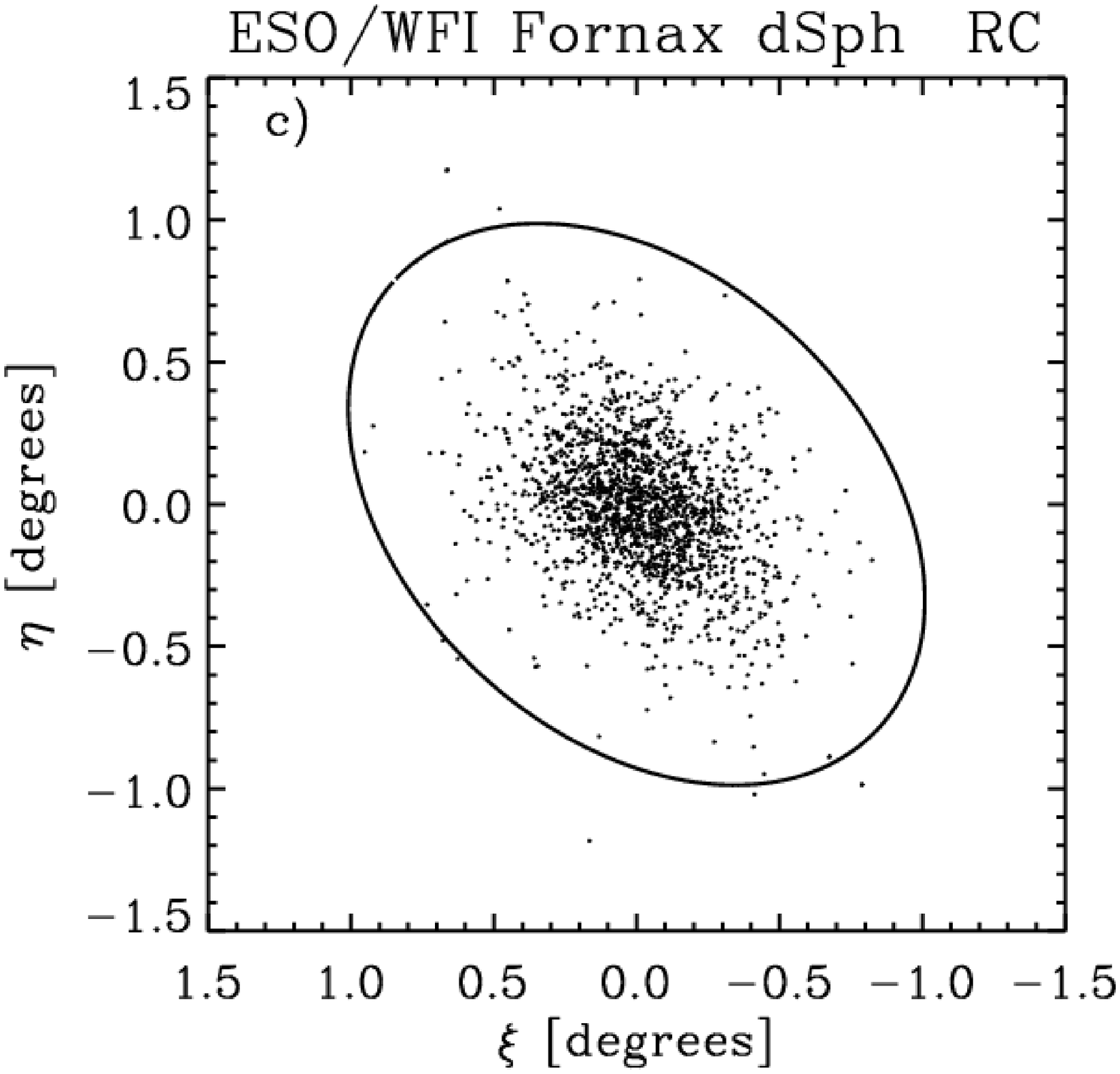}
\caption{Spatial distribution of BHB, RHB and RC stars in the Fornax dSph as selected from the CMD shown in 
Fig.~\ref{fig:cmd}b. The RC stars show a more concentrated 
and less extended distribution than RHB and BHB stars. The number of RC stars has been normalised 
to the number of RHB stars.}
\label{fig:fov_old}
\end{center}
\end{figure}

\begin{figure}
\begin{center}
\includegraphics[width=0.4\textwidth]{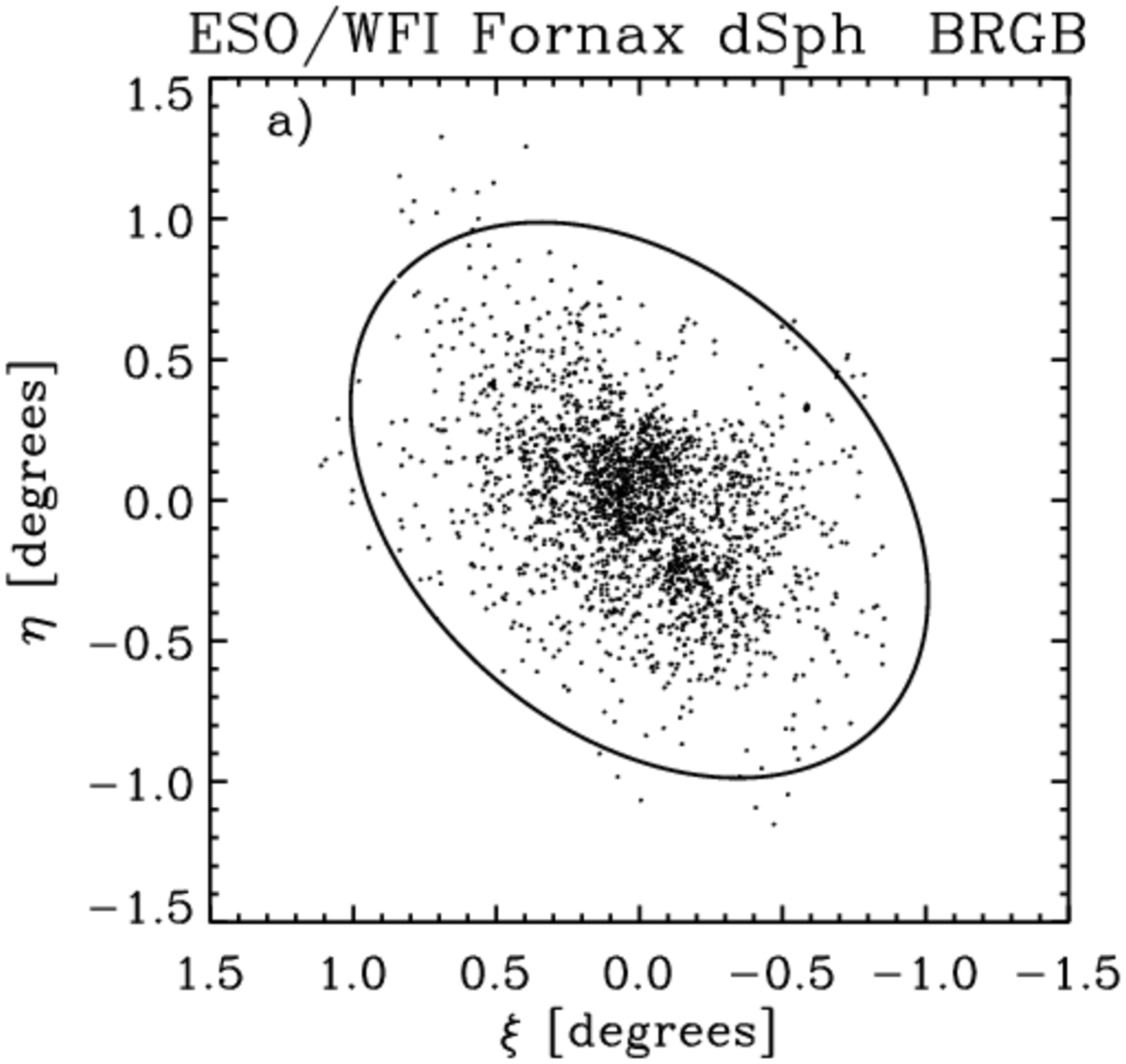}
\includegraphics[width=0.4\textwidth]{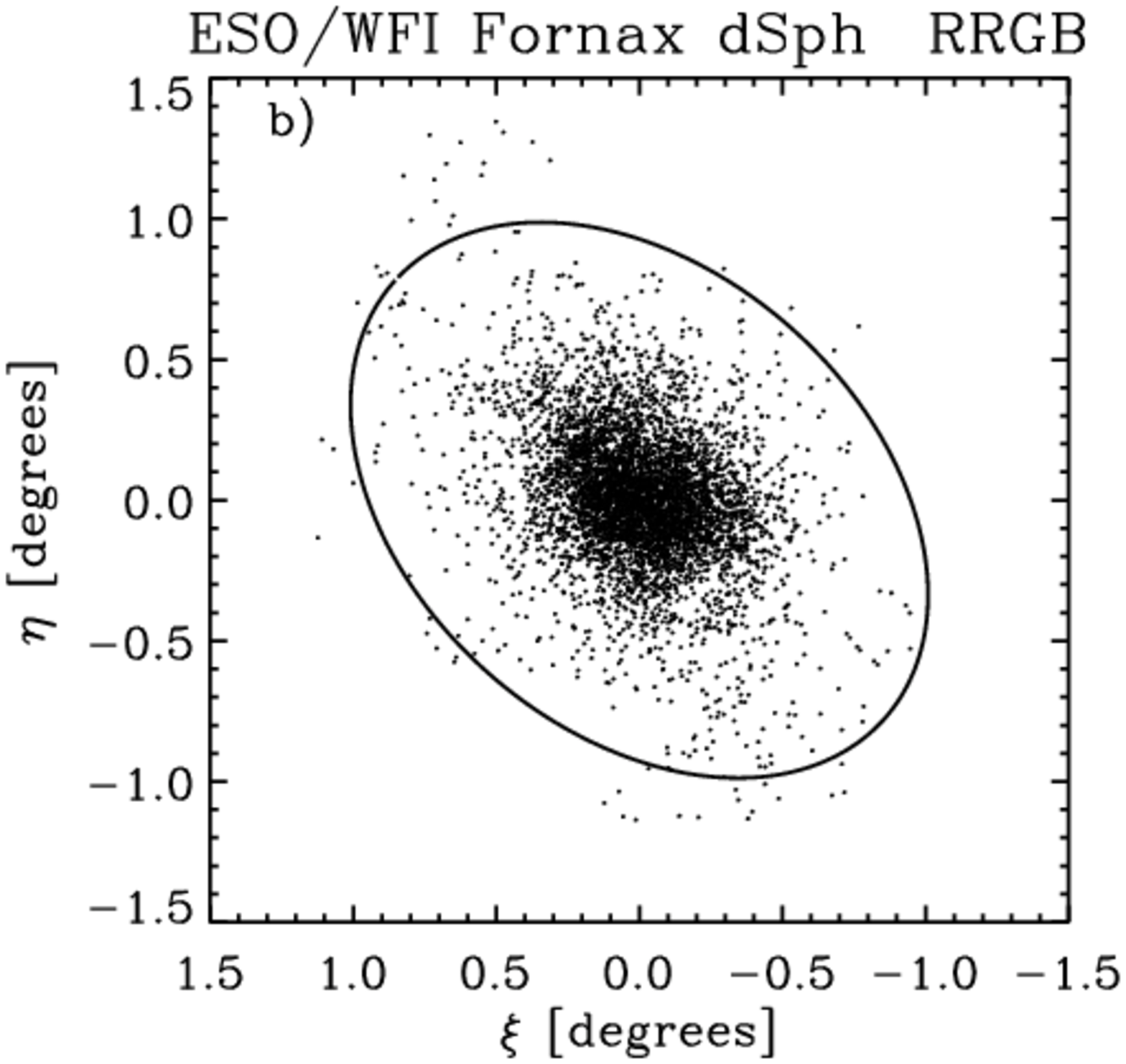}
\caption{Spatial distribution of Blue RGB and Red RGB stars in the Fornax dSph 
(as selected from the CMD shown in Fig.~\ref{fig:cmd}b). }
\label{fig:fov_rgb}
\end{center}
\end{figure}

\begin{figure}
\begin{center}
\includegraphics[width=110mm]{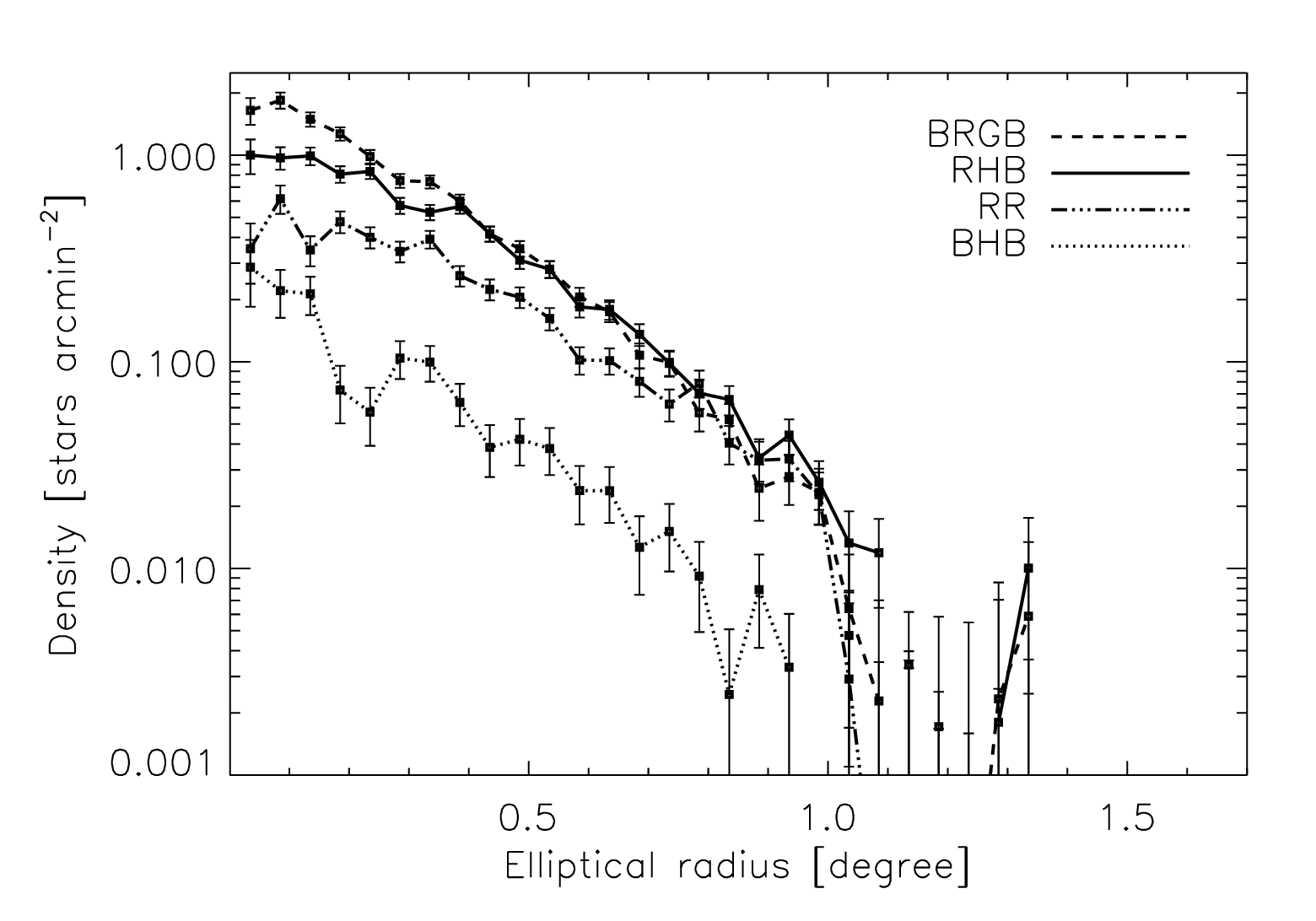}
\includegraphics[width=110mm]{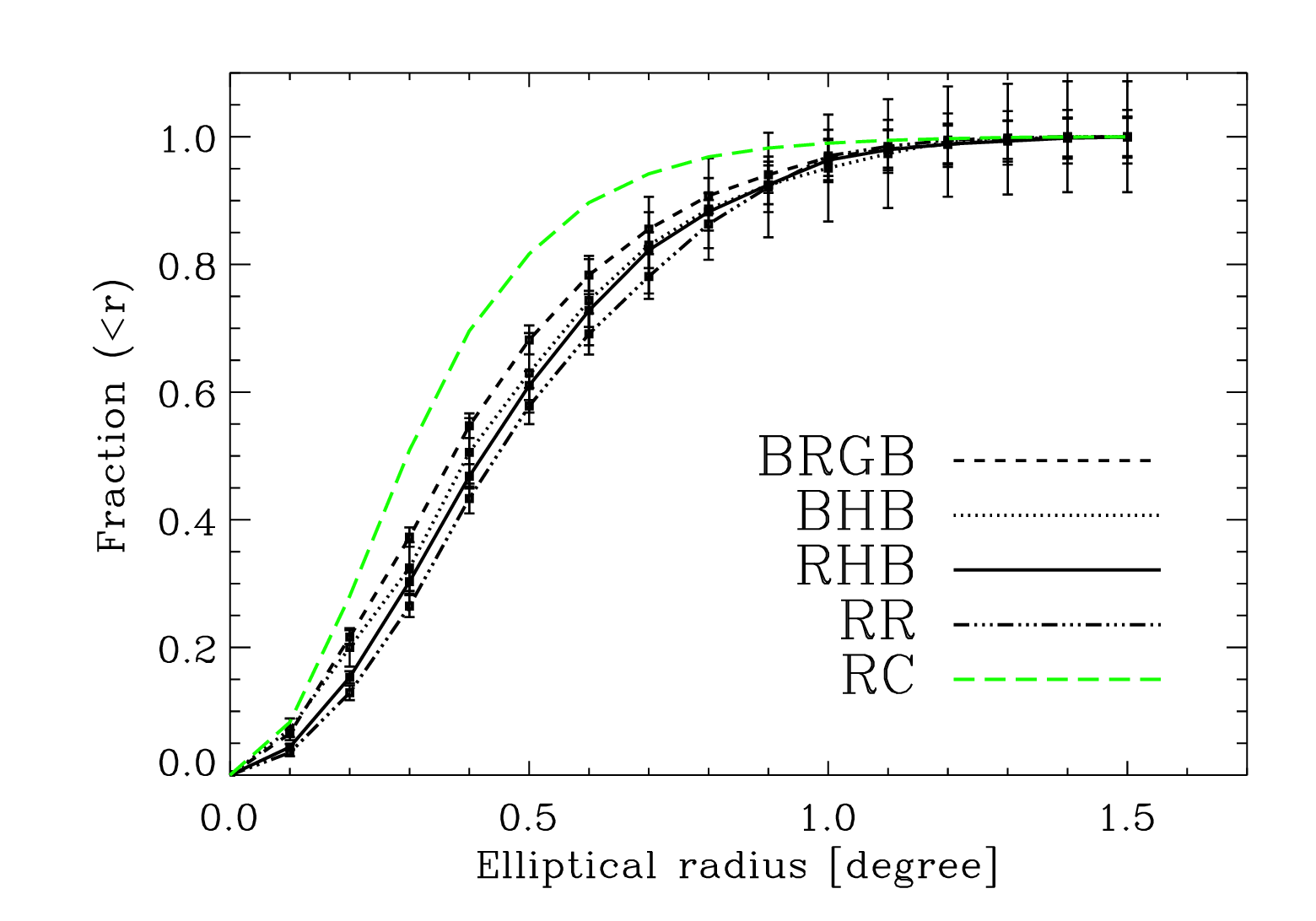}
\caption{Surface number density (top) and fraction of stars within elliptical radius $r$ (bottom) 
of {\bf ancient} stellar populations in the Fornax dSph (solid line: RHB stars; dotted line: BHB stars; 
dash-dot-dot line: RRLyrae stars; dashed line: B-RGB stars). 
The surface number density has been Galactic stellar contamination subtracted. Note that all the 
ancient stars, even if found at different stages of stellar evolution, 
display a similar spatial trend with radius.    
In the bottom panel we show for comparison the fraction of stars within elliptical radius $r$ for RC stars 
(long dashed line), 
representative of intermediate age stars: note the more extended and less concentrated spatial distribution 
of ancient stars with respect to intermediate age stars. 
}
\label{fig:cum_old}
\end{center}
\end{figure}

In Fornax the BHB stars overlap the MS in the inner regions, 
thus we reduce the selection box of BHB stars to V$-$I$\sim$0.15 - V$-$I$\sim$0.35.
In Fig.~\ref{fig:fov_old}a-b we plot the spatial distribution of the BHB and RHB 
components, and in Fig.~\ref{fig:fov_rgb}a  the location of B-RGB stars. 
Figure~\ref{fig:cum_old} shows the Galactic contamination subtracted 
density profile for the ancient populations (top) and the fraction of stars 
within a radius $r$, $f(<r)$, in bins 
of 0.1 deg (bottom). The ancient stellar components clearly all have the same distribution, 
the only exception being the trend at $r<$0.3 deg for the BHB population, which  
suffers from small number statistics. 
It is also likely that the BHB behaviour is altered by the 
presence in the selection box of MS stars (see Fig.~\ref{fig:cmd}b).
 A $\chi^2$ two-sample test applied to the percentage of stars within a radius $r$ 
out to the estimated Fornax tidal radius gives a 
probability of 99.99\% that the ancient stellar populations are drawn 
from the same distribution.
 
Taking as reference the RHB cumulative distribution, we can see that 
95\% of the ancient population is found within $r\lesssim$ 0.9 deg. 
The dominant, 
intermediate age, stellar population in Fornax, i.e. RC stars,  
shows a distinctly different distribution, 
with 95\% of its stars distributed within $r\lesssim$ 0.7 deg. 
Clearly, the ancient 
populations display a more extended spatial distribution then the intermediate age stars. 
The same $\chi^2$ test applied to the RHB and RC stars 
gives a low probability, 5\%, that these two populations are drawn from 
the same parent population, confirming the fact that intermediate and 
ancient stars show very different spatial distributions.

There is evidence for small scale spatial substructure in part of the ancient 
stellar component of Fornax. 
In Fig.~\ref{fig:fov_rgb}a (B-RGB) there is a clear overdensity of stars 
at $\xi \sim -0.15$ and  $\eta \sim -0.25$ ($\alpha \sim 2^h 39^m 8^s$, 
$\delta \sim -34^{\circ} 45^{'} 51^{``}$); 
this also appears to be present, though 
less pronounced, in the 
spatial distribution of RHB stars at similar position, however in 
contrast the 
distributions of BHB and RRLyrae stars appear to be smooth. 
We find the number of B-RGB within 4$'$ 
of the above feature significantly larger (within 2$\sigma$) with respect 
to the number within 3 control fields, of the same dimensions and at the same elliptical radius; 
the overdensity of RHB is instead not statistically significant.
A CMD centred on this overdensity shows 
a slightly bluer colour for the RGB in the over-density than in 3 control fields. 

There is no 
evidence that the nearby cluster 2 ($\xi \sim -0.23$, $\eta \sim -0.30$ in Fig.~1) is 
related to this over-density, because the stars in the over-density do not match 
the cluster 2 CMD \citep[e.g.][]{buonanno1998}. 

Similar overdensities in the stellar distribution have been found in previous works both 
in other dSphs as well as in Fornax dSph (e.g. Coleman et al. 2004).

\subsubsection{Intermediate age populations}
\begin{figure}
\begin{center}
\includegraphics[width=100mm]{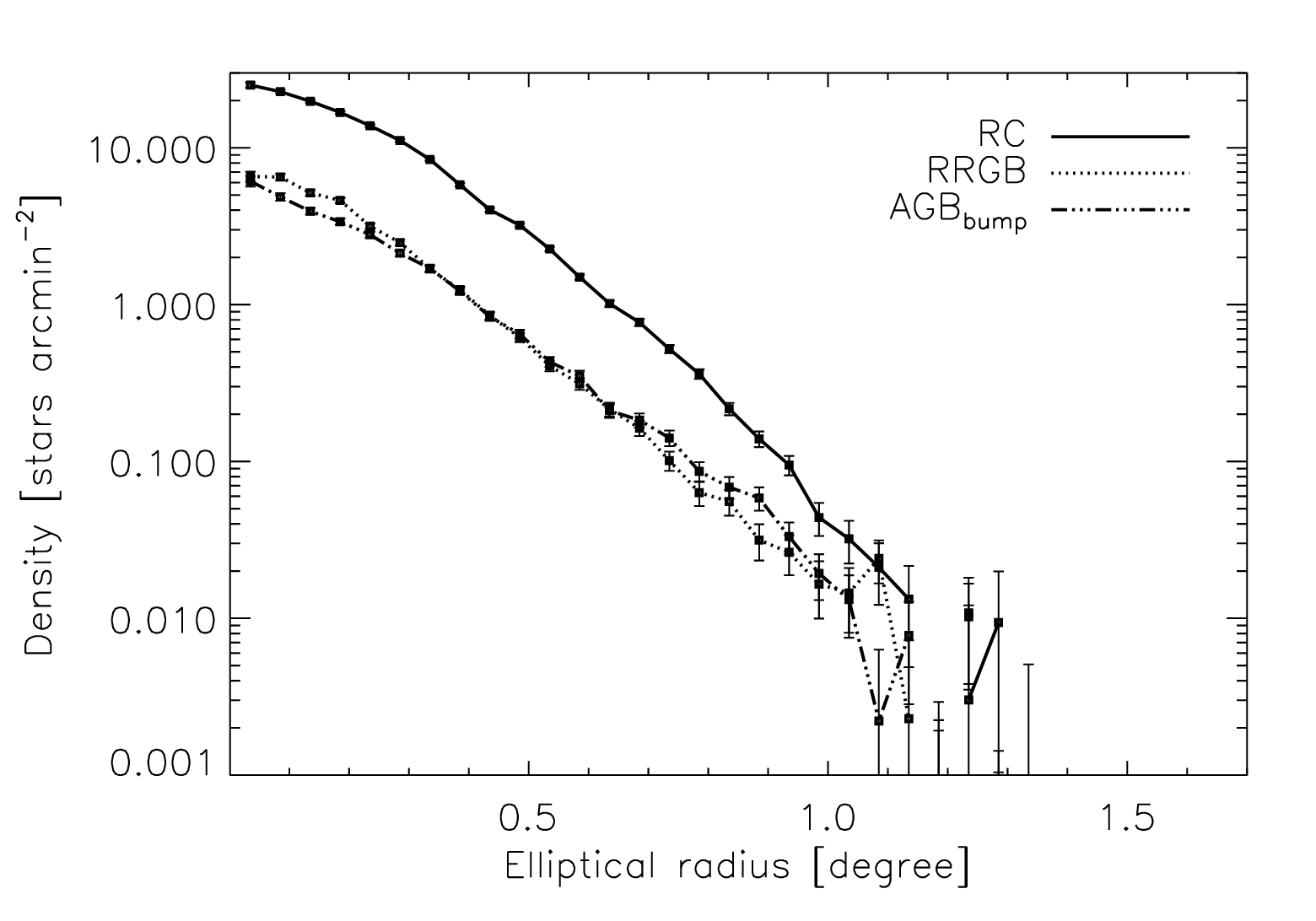}
\includegraphics[width=110mm]{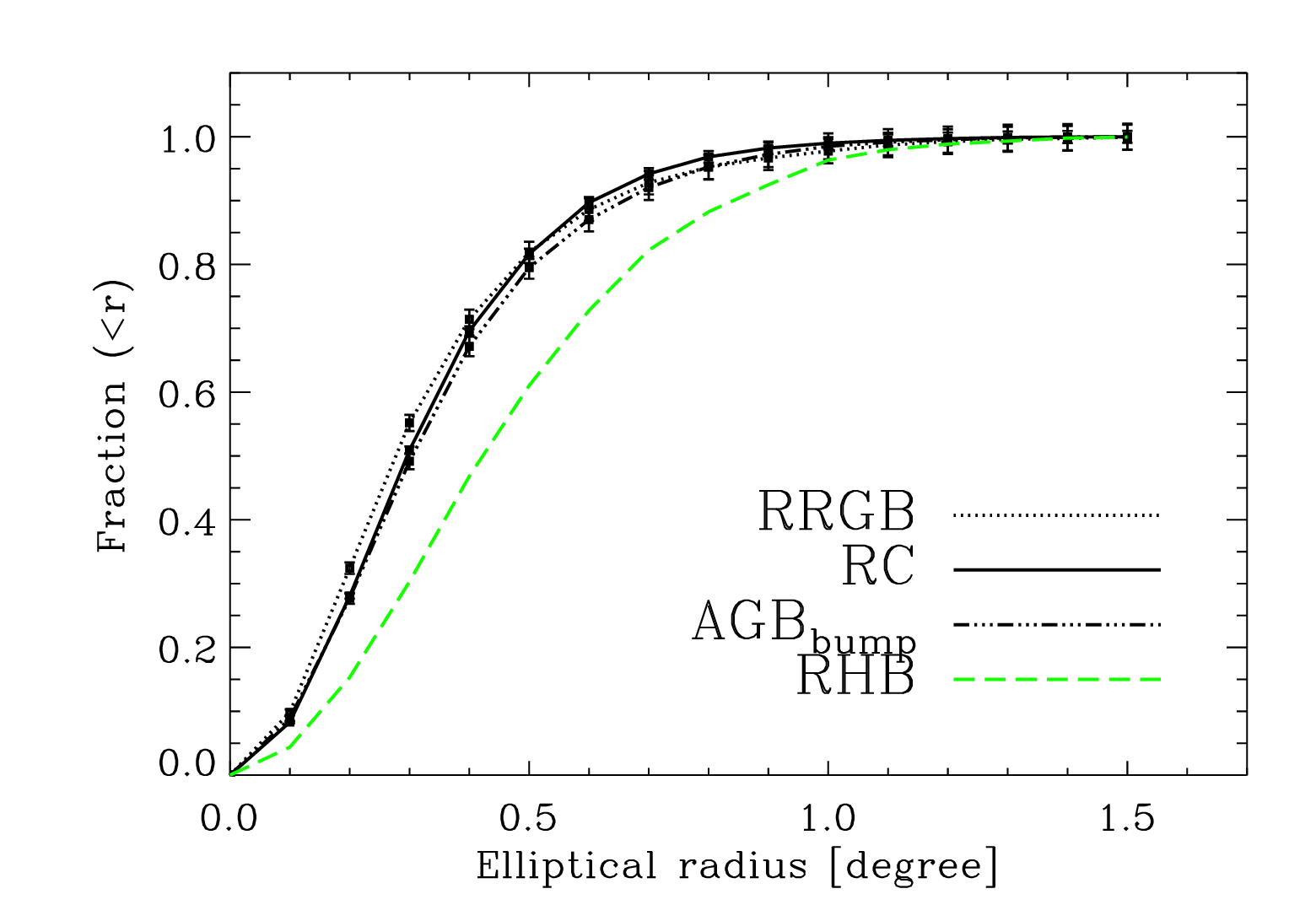}
\caption{
As in Fig.~\ref{fig:cum_old} but for {\bf intermediate age} stellar populations (solid line: RC stars; 
dotted line: R-RGB stars; dash-dot-dot line: AGB bump stars). All the 
intermediate age populations 
display a similar spatial trend with radius.
In the bottom panel we show for comparison the fraction of stars within elliptical radius $r$ for RHB stars 
(long dashed line), 
representative of ancient stars, as in Fig.~9: note the more extended and less concentrated spatial distribution 
of ancient stars with respect to intermediate age stars.
}
\label{fig:cum_int}
\end{center}
\end{figure}

The spatial distribution of the intermediate age populations 
(2-8 Gyr old) is represented by 
RC, AGB bump and R-RGB stars distribution. Figure~\ref{fig:fov_old}c shows the location 
on the field of view of the selected RC stars. As already mentioned, 
the RC stars form the dominant stellar population in Fornax. 
Thus, to allow a more meaningful comparison between the spatial distributions of 
intermediate age and ancient stars, we normalised 
the RC population to the number of stars present in the RHB. 
Visual comparison of these two distributions shows that RC stars have a less 
extended and more concentrated spatial distribution than RHB stars. The distribution 
of R-RGB stars (Fig.~\ref{fig:fov_rgb}b) is similar to the RC stars, indicating 
a similar origin and thus similar age/metallicity properties for the stars 
found in the red part of the RGB in Fornax and in the RC.
 Figure~\ref{fig:cum_int} shows the Galactic contamination subtracted 
density profile (top) and the cumulative fraction (bottom) of RC, AGB bump and 
R-RGB: they clearly have the same spatial distribution; 
a $\chi^2$ test applied to the percentage of stars within a radius $r$ 
out to the estimated Fornax tidal radius gives a 
probability of 99.99\% that the intermediate age stellar populations are drawn 
from the same distribution.
Thus, it appears likely that all the stars with ages of 2-8 Gyr formed from the same gas distribution.

\subsubsection{Young populations}

\begin{figure}
\begin{center}
\includegraphics[width=0.47\textwidth]{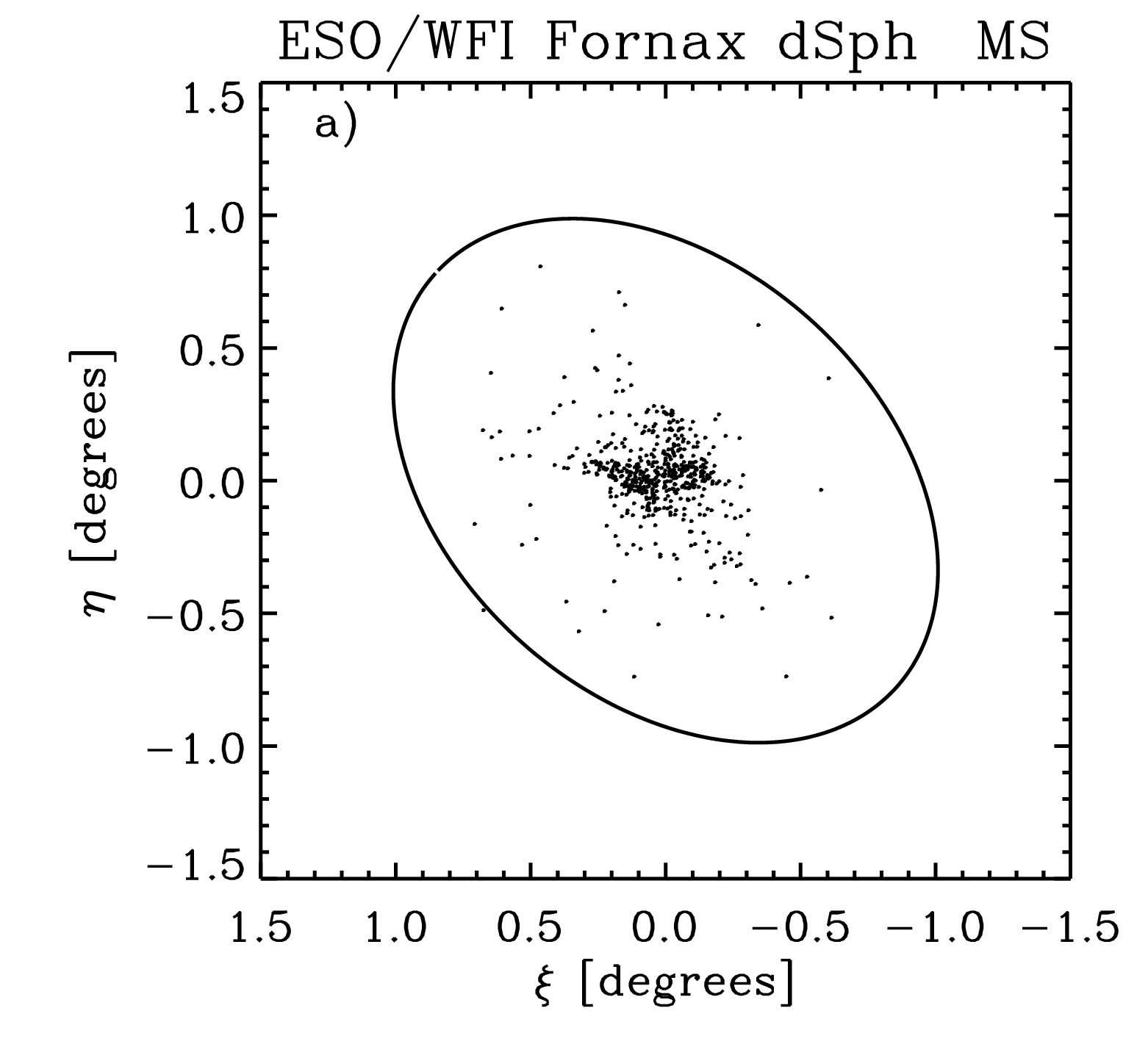}
\includegraphics[width=0.47\textwidth]{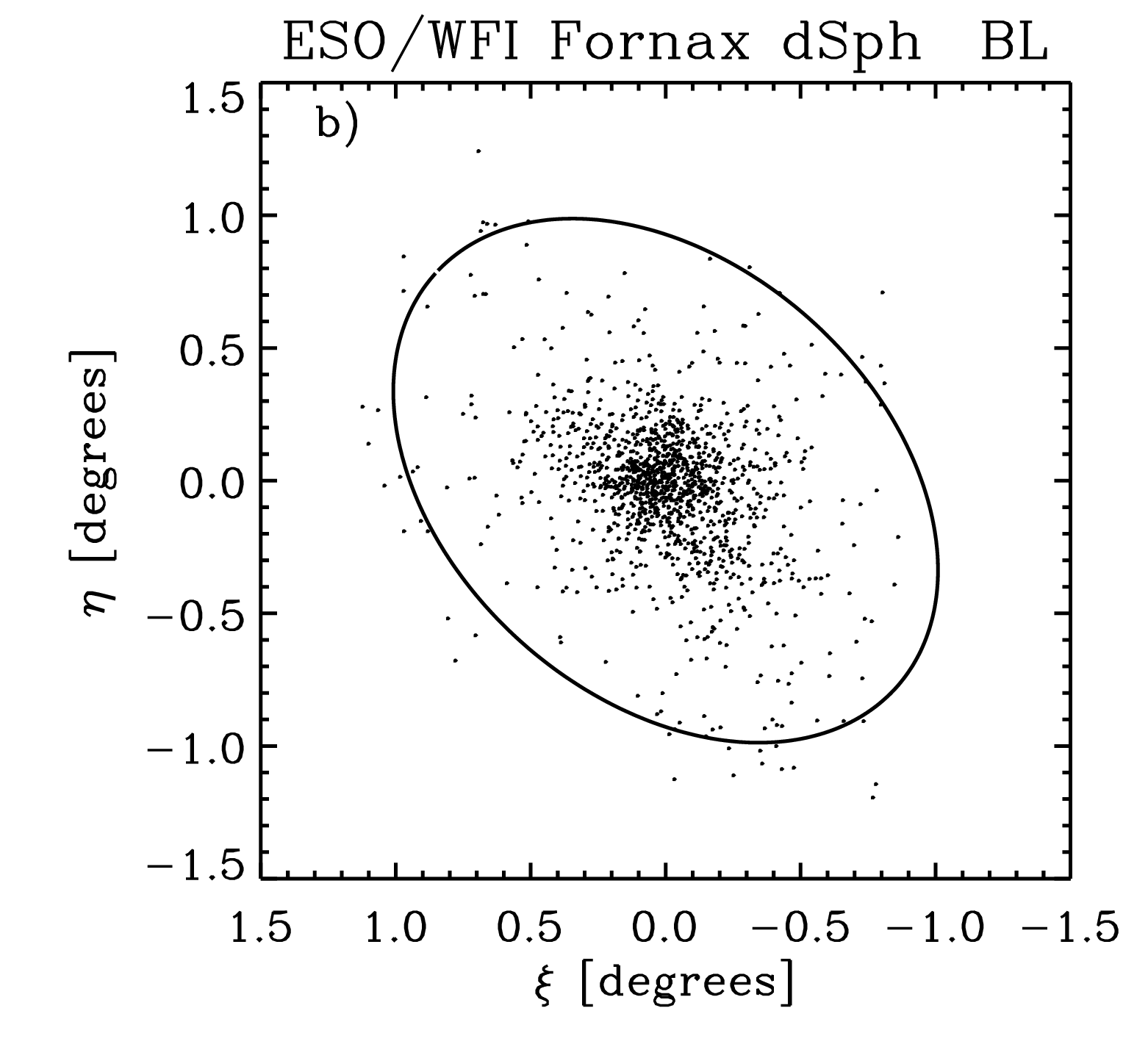}
\includegraphics[width=0.47\textwidth]{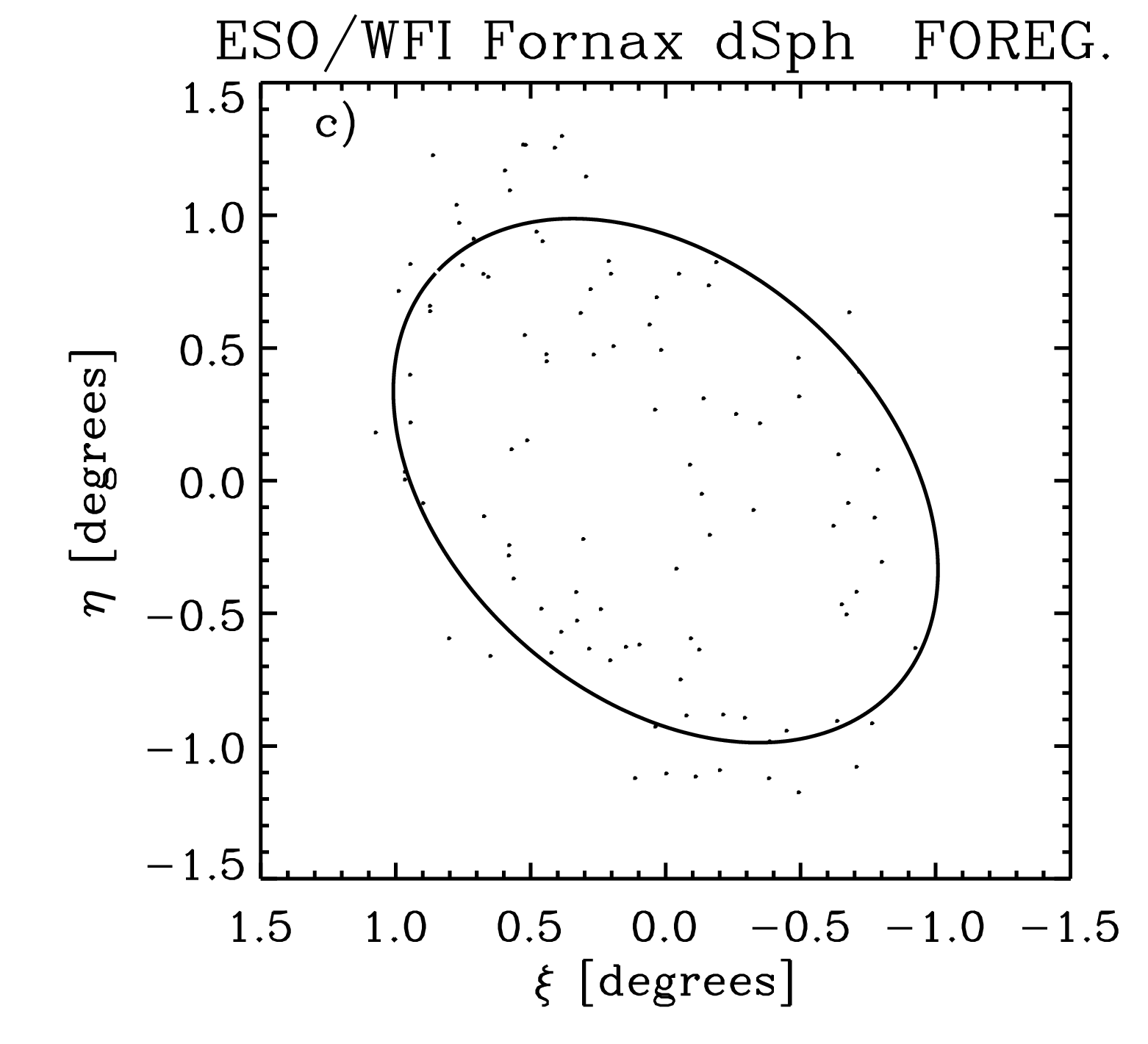} 
\caption{Spatial distribution of main sequence (a) and Blue~Loop (b) stars 
in the Fornax dSph (as selected from the CMD shown in Fig.~\ref{fig:cmd}b) and c) 
the foreground contamination in the Blue~Loop distribution as from 
a CMD-selected sample of foreground stars to match the Blue~Loop contamination density. 
Note the asymmetric 
distribution of MS stars in a), 
elongated in the E-W direction and tilted by $\sim 40^{\circ}$ with respect 
to the main body of Fornax dSph. The inner shell-like feature detected by Coleman et al. (2004) 
is located at $\xi \sim$0.12, $\eta \sim -$0.20 and elliptical radius of 0.33. This is also 
the position of one of the two fields observed by Olszewski et al. (2006).}
\label{fig:fov_young}
\end{center}
\end{figure}

We have already seen from the CMDs in Fig.~\ref{fig:cmd_r} that young MS ($<$ 1 Gyr old) and BL stars 
are only present within $r<$ 0.4 deg, in the central region of Fornax. 
Figure~\ref{fig:fov_young} shows the spatial distribution of MS and Blue~Loop 
stars (panel a,b). 
The main sequence stars are 
centrally concentrated and show a different, asymmetric distribution with respect 
to the older stars. They predominantly form a structure extended 
in the E-W direction, with a major axis 
tilted at $\sim 40^{\circ}$ with respect to the bulk of the stellar population (as already 
noted by Stetson et al. 1998). The overall spatial distribution of BL stars 
is also asymmetric, although not as pronounced as the MS. 
It is difficult to remove the effects of the foreground contamination (Fig.~\ref{fig:fov_young}c) 
but it appears that the blue
loop stars have a more relaxed distribution than the MS.

\begin{figure}
\begin{center}
\includegraphics[width=100mm]{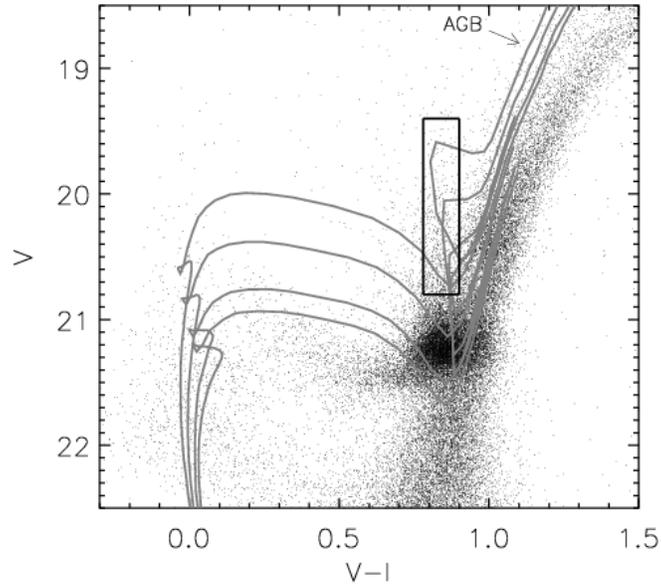}
\caption{CMD for stars within 0.4 degree of the 
centre of the Fornax dSph; the box shows the selection region for the Blue~Loop 
stars. The colours and magnitudes of the Blue~Loop 
stars are consistent with theoretical Padua isochrones (Girardi et al. 2000) 
of metallicity Z=0.004 ([Fe/H]$\sim-$0.7) and 
age 0.4,0.5,0.6, 0.7 Gyr (from top to bottom). The stars forming the BL feature 
have mass in the range 2.3- 2.8 \sm. 
}
\label{fig:cmd_loop}
\end{center}
\end{figure}

As mentioned in Sect.~3.2.3 the colour of the BL feature is 
strongly dependent on metallicity. Overlaying 
Padua stellar isochrones \citep{girardi2000} 
on the CMD (Fig.~\ref{fig:cmd_loop}), 
we find that the Blue~Loop stars are consistent
with ages between 400 and 700 Myr and metallicity Z=0.004.  
Isochrones of lower metallicities (Z=0.001) predict 
the presence of a plume of Blue Loop giants much closer to the MS than observed, 
and the predicted average colour of the BL feature 
for higher metallicities (Z=0.008) is significantly 
redder than observed\footnote{Using different sets of isochrones, e.g. 
Geneva \citep{Lejeune2001} 
or Yonsei-Yale \citep{yi2001, kim2002}, would bring to slightly different results, however qualitatively in 
agreement.}.

Figure~\ref{fig:cmd_loop} also shows that the Girardi et al. (2000) Padua isochrones 
suggest that the 
MS contains stars as young as 200-300 Myr,
 and as old as 1 Gyr, which is different with respect to the BL stars age range. 
Thus, to compare the spatial distribution of MS and BL stars,  
we must separate the MS stars in two groups, those predominantly younger and 
those older than 400 Myr. This is 
achieved by splitting the MS in colour: the ``young'' and ``old'' MS stars are approximately 
bluer and redder, respectively, than $V-I=$0. In this case we find that the spatial distribution 
of ``old'' MS stars is consistent with the BL spatial distribution, whilst ``young'' MS 
stars display a more asymmetric and centrally concentrated distribution. 
Thus, it seems that the ``young'' MS stars are found closer to the location 
where they were formed, whilst ``old'' MS and BL stars have diffused 
to larger radii. To estimate the diffusion time, we use the dynamical time 
(Binney \& Tremaine 1987), which ranges between 350- 800 Myr, 
assuming a mass within the tidal radius between 
10$^8$-1.55$\times 10^7$ \sm. This is consistent with the age of the ``old'' MS stars. 
Thus the elongated structure of the MS is formed predominantly by the younger MS stars, and is 
presumably the remaining signature of the collapsed gas from which the stars were formed.

\subsubsection{A quantitative comparison between stellar populations}
We have shown that the younger the stellar population the more 
concentrated and less extended is its spatial distribution over all the observed 
populations. We have also shown 
that stars in the same age range (but different stellar evolutionary phases) 
display very similar spatial distributions, 
and that the blue and red side of the RGB are linked to the ancient 
and intermediate age populations, respectively. 
A $\chi^2$ test is applied to the percentage of stars within a radius $r$ 
out to 
the estimated Fornax tidal radius for the several 
stellar components (Figs.~\ref{fig:cum_old}, \ref{fig:cum_int}) 
gives a probability of 99.99\% that RC, AGB and R-RGB stars are drawn 
from the same distribution, and again 99.99\% for the different 
ancient stellar populations, whilst  
when comparing the distribution of an ancient population, e.g. RHB, 
to one of the intermediate age population, e.g. RC, we obtain a probability of 5\% 
that they are drawn from the same spatial distribution. 
This shows that we can consider stars found at different stages of stellar evolution 
but with similar ages as belonging to the same population. 

We derived density profiles, from which Galactic contamination has been subtracted,  
for these two components and compared them to the models used in Sec.~3.1.
The best-fitting profile for the old population is given 
by the King model, that follows more closely the data beyond 
1 deg with respect to the Sersic profile. However, the density profile of the 
intermediate age stars is better represented by a Sersic profile, 
since the King model declines too steeply. In both cases the exponential and 
Plummer models overpredict the density for $r\lesssim$ 0.2 deg and $r\gtrsim$ 0.9 deg, 
and underpredict it at intermediate radii. 
Independently from the adopted model, 
the results of the fit (see Table~\ref{tab:par_oldint}) show that intermediate age stars are more 
centrally concentrated and less extended than old stars.

The spatial distribution of different stellar populations 
is a powerful way to identify stellar populations
of different ages in a dwarf galaxy.

\subsection{Shells}

The discovery of two shell-like features
was recently reported in Fornax from photometric studies 
\citep{coleman2004, coleman2005}. The first of these features 
was identified outside the core radius at $\alpha= 2^h 40^m 28.5^s$ and 
$\delta= -34^{\circ} 42' 33''$ (corresponding to 
$\xi \sim$0.12, $\eta \sim -$0.20 and elliptical radius of 0.33 deg in 
Fig.~\ref{fig:coverage}) as a small over-density of stars in data 
down to $I \sim$21 \citep{coleman2004}. Adding 
deeper data from Stetson et al. (1998) to their sample the authors found the shell 
to be consistent with being dominated by young, 2 Gyr old, ZAMS stars.
The authors proposed that this might be analogous to the 
shells found in elliptical galaxies and suggested that Fornax might have accreted a smaller 
gas-rich galaxy around 2 Gyr ago. 

A second ``shell'' was found in the RGB distribution outside 
the tidal radius, around 1.3$^{\circ}$ northwest from the centre \citep{coleman2005} and, 
similarly to the inner shell,  
located along the minor axis and elongated along the major axis.

\citet{ol2006} carried out much deeper imaging of this over-density, down to $R\sim$ 26 and 
refined the age of the ``feature'' and suggested that the excess of stars 
in the shell might be the result of a burst of star formation 
which occurred 1.4 Gyr ago with a metallicity Z=0.004 ([Fe/H]$=-$0.7), favouring a 
scenario in which the gas 
from which these stars formed was pre-enriched within Fornax itself.

Our photometry covers the region of the inner shell,  
and we do not detect such a feature, but comparing our sample with the additional 
data-set from \citet{stetson1998}, we find signs of the shell when we include the stars 
we had discarded from our analysis because they were detected in $V$ but not in the $I$ band. 
This makes it clear that the feature is just at the detection limit of our data. We also note 
that the MS stars are asymmetrically 
distributed, and this could cause the presence of local overdensities 
with respect to other regions in Fornax, at the same elliptical radius, 
where young stars might be under-represented.
Deeper photometry covering the entire region at $r<$0.4-0.5 deg would make it possible 
to resolve this issue.

\section{Results: Spectroscopy}

\begin{figure}
\begin{center}
\includegraphics[width=120mm]{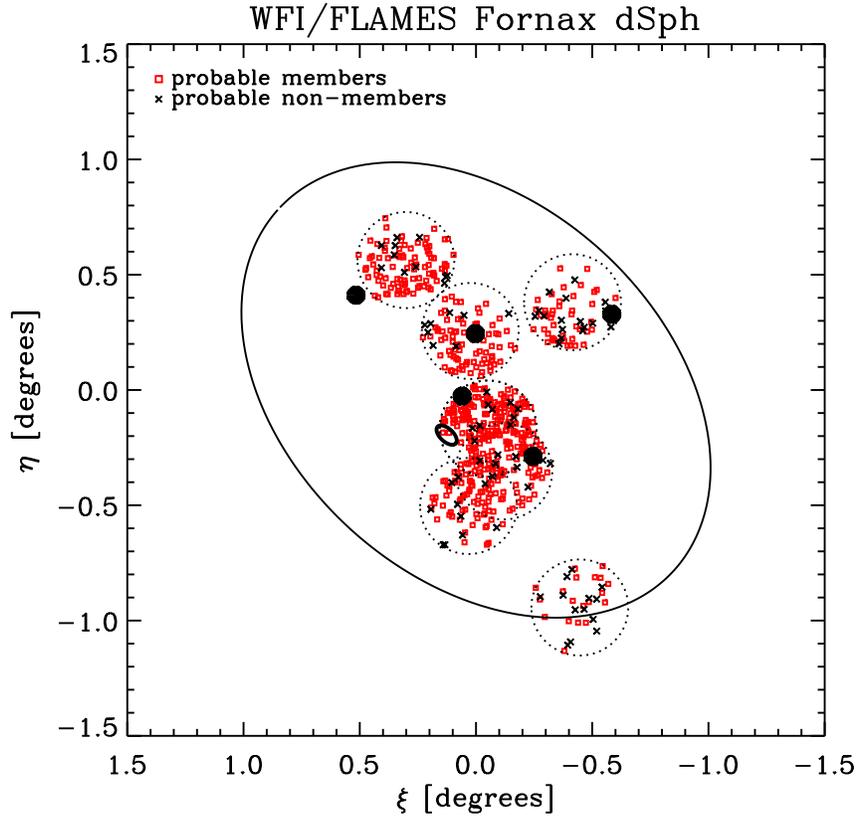}
\caption{Location of the observed FLAMES fields and 
targets (with S/N per \AA\ $>$ 10 and error in velocity $<$ 5 \kms) 
in the Fornax dSph (squares: probable members; 
crosses: probable non-members). 
The black filled circles show the location of Fornax globular clusters, the black small 
ellipse shows the location of the ``shell-like'' feature detected by Coleman et al.~2004, and 
the dashed circles indicate the observed FLAMES fields. The tidal radius is from this work.}
\label{fig:fov}
\end{center}
\end{figure}

Our VLT/FLAMES spectroscopic 
survey of individual stars in the Fornax dSph allows us 
to determine velocities and metallicities 
([Fe/H]) from the CaT equivalent width 
for a subsample of RGB stars out to the nominal tidal radius of Fornax dSph (see Fig.~\ref{fig:fov}).

The dependence of CaT equivalent width on metallicity has been 
empirically proven by extensive work in the literature 
\citep[e.g.][]{arm1988, ol1991, arm1991}. \citet{rutledge1997a} 
presented the largest compilation of CaT EW measurements 
for individual RGB stars in globular clusters, which \citet{rutledge1997b} 
calibrated with high resolution metallicities, proving the CaT method to be 
reliable and accurate in the probed metallicity range, $-2.2 <$[Fe/H]$<-0.6$.  

We obtained metallicities from the CaT equivalent widths by using the relation 
derived by \citet{rutledge1997b}:
\begin{equation}
{\rm [Fe/H]}= -2.66 + 0.42\, W'
\end{equation}
For deriving the reduced equivalenth width $W'$ we used the equation below, as in Tolstoy et al. (2001):
\begin{equation}
W' = (W_{\rm 8542} + W_{8662}) + 0.64 (V - V_{\rm HB})
\end{equation}
where $W_{\rm 8542}$ and $W_{8662}$ are the equivalenth width 
of the two strongest CaT lines, at $\lambda= 8542.09$ \AA\ and $\lambda= 8662.14$ \AA\ ; 
$V$ is the apparent $V$ magnitude of the star; and $V_{\rm HB}$ the 
mean magnitude of the horizontal branch. We checked this calibration against 
VLT/FLAMES HR measurements of an overlapping sample of 90 stars 
(Hill et al. 2006) to confirm our calibration and accuracy of CaT metallicity determination. 
Details will be provided in Battaglia et al. (2006b).

Since the Fornax dSph displays a very broad RGB (see Sect.~3.2 and Fig.~\ref{fig:cmd}) we chose 
our targets from a box around the RGB covering a wide colour range, 
to avoid biasing our sample in age or in metallicity. The spatial 
distribution of our targeted VLT/FLAMES is shown in Fig.~\ref{fig:fov}, 
together with the position of Fornax globular clusters 
and the location of the shell-like feature \citep{coleman2004}. 
The fields cover a wide range of radii. Since, as discussed in Sect.~3, 
the stellar population of different ages display different spatial 
distributions, our sample will contain predominantly intermediate age 
stars in the central fields, and ancient stars in the outermost fields. 
Previous studies, which were restricted to the central regions, 
were clearly biased towards the properties of the intermediate age population.  

\begin{figure}
\begin{center}
\includegraphics[width=80mm]{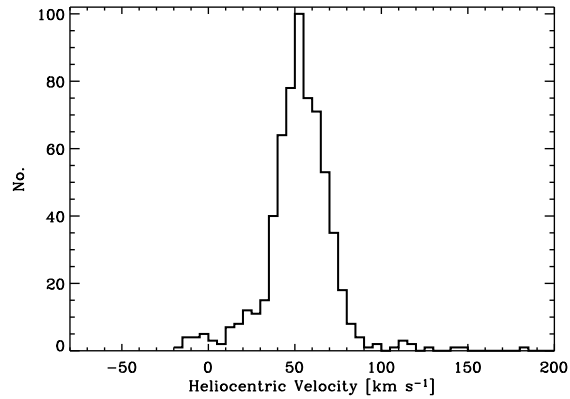}
\caption{Histogram of velocities for the observed VLT/FLAMES targets in the Fornax dSph 
which passed our S/N and velocity error criteria (641 stars).}
\label{fig:histo_vel}
\end{center}
\end{figure}

Figure~\ref{fig:histo_vel} shows the velocity histogram 
of our VLT/FLAMES targets which met our S/N and velocity 
errors criteria (641 stars). To find the systemic velocity of Fornax we first selected 
the stars found within 4$\sigma$ of the velocity 
peak associated with Fornax. We used the  
value of $\sigma=$ 13 \kms 
taken from the literature (e.g. Walker et al. 2005) as a first approximation. 
We calculated the mean velocity and the velocity dispersion from the 4$\sigma$ sample and then 
repeated the procedure selecting stars within 3$\sigma$ and finally the stars within 
2.5 $\sigma$. We found a systemic velocity $V_{\rm sys}=53.1\pm$0.6 \kms and 
a velocity dispersion $\sigma_{\rm los}= 13.7 \pm$0.4 \kms (3$\sigma$), and 
$V_{\rm sys}=54.1\pm$0.5 \kms, $\sigma_{\rm los}= 11.4 \pm$0.4 \kms (2.5$\sigma$). 

Because Fornax has a systemic velocity which strongly overlaps with the Galaxy, 
we tested to see if the large variation in velocity dispersion for different $\sigma$ cuts 
might be due to 
foreground contamination. We perform Monte Carlo simulations 
producing 10000 sets of 600 velocities about the mean velocity of Fornax  
Gaussianly distributed with a trial dispersion 
of 15 \kms. The dispersions obtained cutting the artificial velocity 
set at 3 $\sigma$ and 2.5 $\sigma$ were consistent within 2 $\sigma$ in 
the 99.63\% of the cases. We obtained very similar results when we fixed 
the dispersion at 12 \kms, and/or convolved the simulated velocities 
with observational errors. 
This means that the differences in velocity dispersion 
we obtain with the $3 \sigma$ and $2.5 \sigma$ cuts for our data are most likely not 
due to the cuts themselves, but to the presence of Galactic 
contamination. 
A similar result is obtained by adding a Galactic component to the simulated velocities of Fornax, 
as a Gaussian with parameters determined from the Besan\c con model \citep{robin2003}. 
The input velocity dispersion for 
the dSph velocity distribution is recovered with the $2.5 \sigma$ cut, 
whilst the dispersion from the $3 \sigma$ cut is overestimated. 
We thus adopt the 2.5$\sigma$ selection about the systemic velocity to limit the foreground 
contamination.

\begin{figure}
\begin{center}
\includegraphics[width=\textwidth]{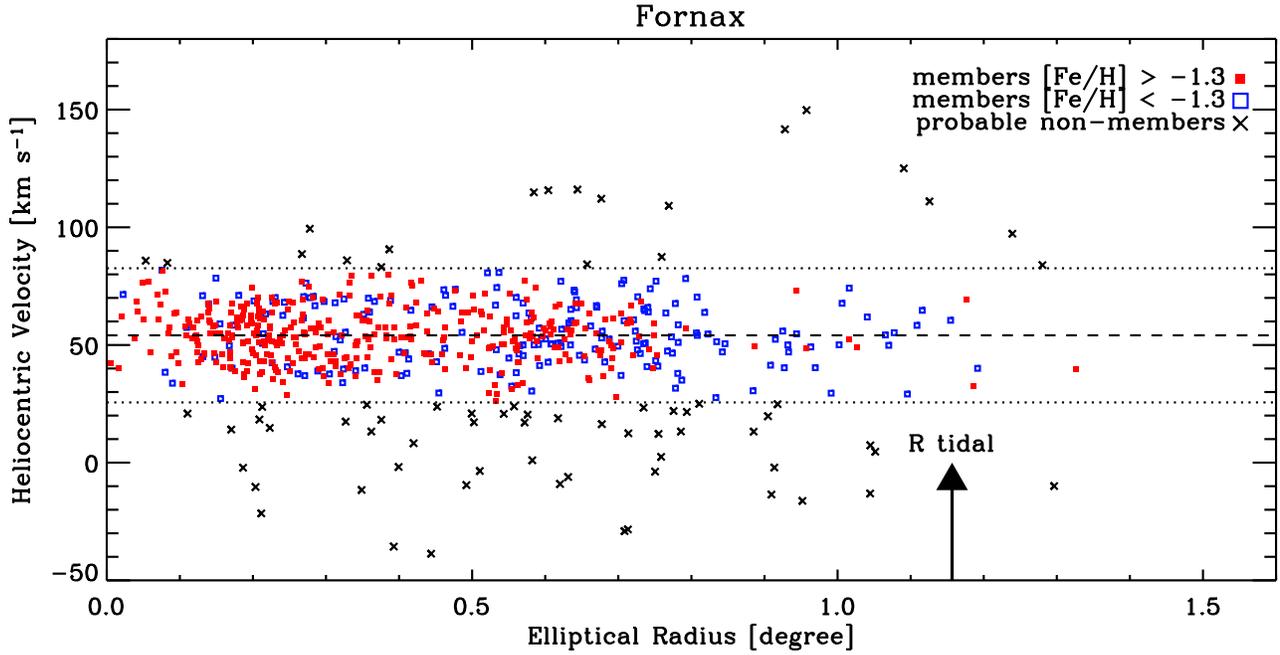}
\caption{Heliocentric velocity with elliptical radius from VLT/FLAMES spectroscopic observations 
of individual RGB stars in the Fornax dSph. We plotted all the stars which satisfy our selection criteria (S/N per \AA\ $>$ 10 and velocity error $<$ 5 \kms). 
The dashed line is the systemic velocity, $V_{\rm sys}= 54.1$ \kms, and the dotted 
lines encompass the 2.5$\sigma$ region of our kinematic selection.
The squares show Fornax velocity members 
(filled: metal rich members, [Fe/H]$>-1.3$; open: metal poor members, [Fe/H]$>-1.3$), and the crosses 
foreground contamination.}
\label{fig:vel}
\end{center}
\end{figure}

\begin{figure}
\begin{center}
\includegraphics[width=80mm]{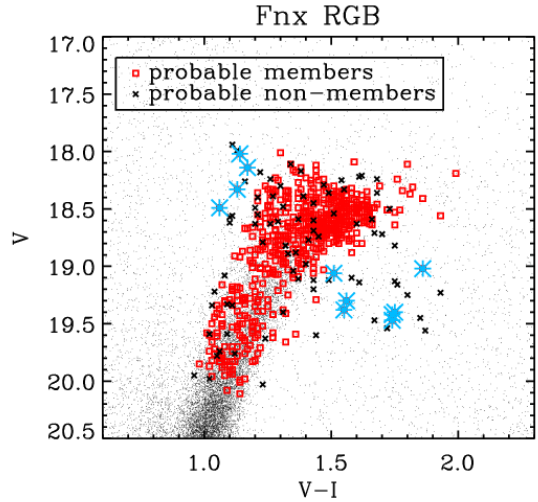}
\caption{Location on the CMD of the observed FLAMES targets for Fornax dSph (squares: probable members; 
crosses: probable non-members). The boundaries of 
our target selection region closely follow the location of the targets. 
The asterisks indicates stars selected as velocity members 
falling in unusual position in the CMD; they are likely to be Galactic foreground contamination 
(see Fig.~\ref{fig:fe_vel_all}).}
\label{fig:cmd_FL}
\end{center}
\end{figure}

\begin{figure}
\begin{center}
\includegraphics[width=110mm]{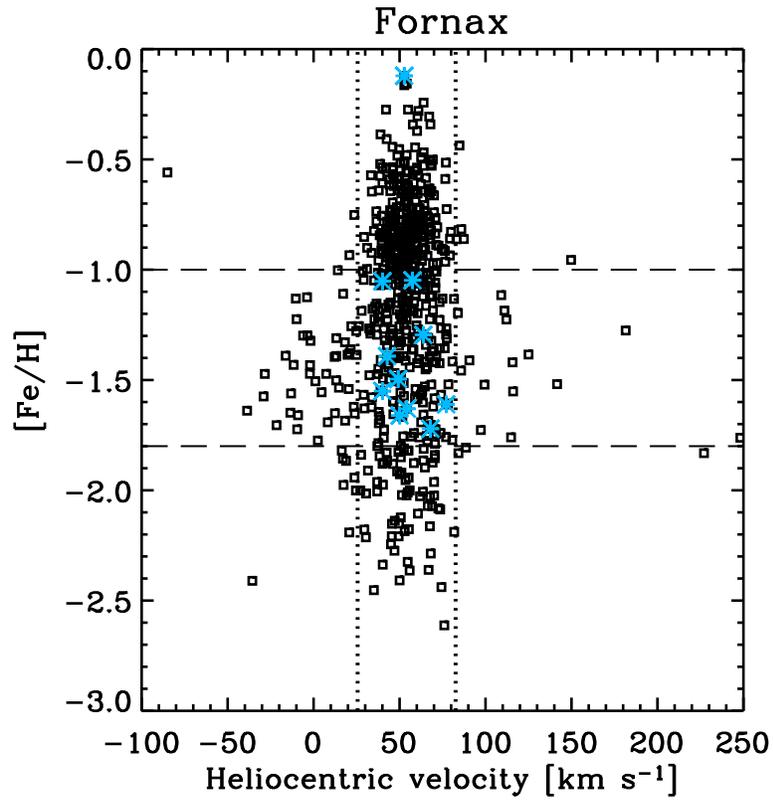}
\caption{Metallicity versus heliocentric velocity for the observed FLAMES targets with 
S/N per \AA\  $>10$, and error in velocity $<$ 5 \kms. The dotted lines show the region of 
our kinematic selection. Note that Galactic contaminants predominantly 
cluster between $-1.8<$[Fe/H]$<-1.0$ (horizontal lines).
 The asterisks show the Fornax velocity members
 present in unusual position in the CMD (see Fig.~\ref{fig:cmd_FL}).}
\label{fig:fe_vel_all}
\end{center}
\end{figure}

From our kinematic selection the number of probable Fornax members 
(with S/N per \AA\  $>$10 and error in 
velocity $<$ 5 \kms) is 562 (see Fig.~\ref{fig:vel}).
 We list in Table~\ref{tab:data_spectro} the relevant informations for the observed 
targets that passed our selection criteria. 
Note that 
probable members are also observed beyond the nominal tidal radius 
(derived in Sect.~3.1). 
In Fig.~\ref{fig:cmd_FL} we show the location on the CMD of probable velocity members and 
non-members. Because of the low heliocentric velocity of Fornax it is 
not always possible to firmly distinguish between members and non-members according to 
a pure kinematic selection: some members are present in a region of the CMD 
where none are expected. Visual inspection of these spectra shows that all but one 
are consistent with a RGB spectrum. However it is difficult to distinguish between 
dwarf and RGB stars in the spectral range covered by our data. 
The distribution of metallicity versus velocity (Fig.~\ref{fig:fe_vel_all}) 
shows that Fornax members are recognizable in this parameter space and the foreground 
stars (kinematic non-members) 
appear to cluster predominantly in the region between $-1.8 <$ [Fe/H] $< -1$. 
The Fornax members off the RGB (asterisks in Fig.~\ref{fig:cmd_FL})
are also identified in Fig.~.\ref{fig:fe_vel_all}. All but one 
fall in the region between $-1.8 <$ [Fe/H] $< -1$, thus the distribution of 
these stars in this parameter space again suggests that they are highly likely to be Galactic contamination.  
As the number of these objects is very small and removing or including them 
in our present analysis will not affect the conclusions, we decided 
to include them in the sample. A more strict membership selection 
will be carried out before analysing the detailed kinematics of Fornax \citep{Battaglia2006a}.

\subsection{Chemo-dynamics}

Several spectroscopic studies of individual stars in the Fornax dSph galaxy 
have shown that it is the most metal rich of the Milky Way dSph 
satellites (except for Sagittarius), with a peak metallicity at [Fe/H]$\sim-1.0$ and a 
large metallicity spread \citep{tolstoy2001, pont2004}. 
 The extent of the metal poor and metal rich tail has changed 
depending mainly on the size and the spatial location 
of the observations. The most 
recent study, consisting of CaII triplet measurements of 117 RGB stars 
\citep{pont2004}, showed that Fornax contains
stars as metal poor as [Fe/H]$\sim-2.0$ and as metal rich as $-0.4$ 
with a peak at [Fe/H]$=-0.9$.

\begin{figure}
\begin{center}
\includegraphics[width=130mm]{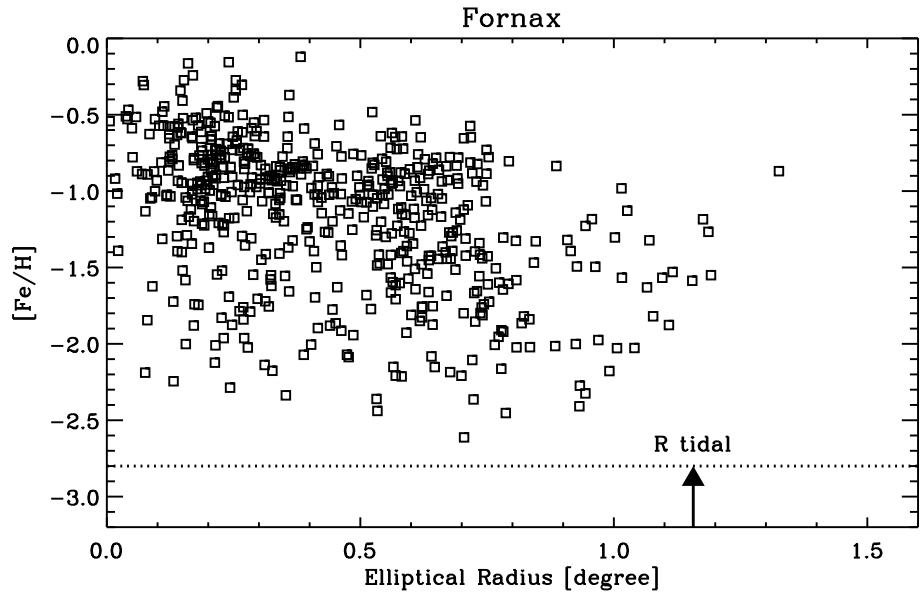}
\caption{Metallicity distribution with elliptical radius for the Fornax dSph. Note the trend 
of decreasing metallicity with radius and the absence of very metal poor stars 
(dotted line indicates [Fe/H]$\lesssim -2.8$).}
\label{fig:met}
\end{center}
\end{figure}

\begin{figure}
\begin{center}
\includegraphics[width=130mm]{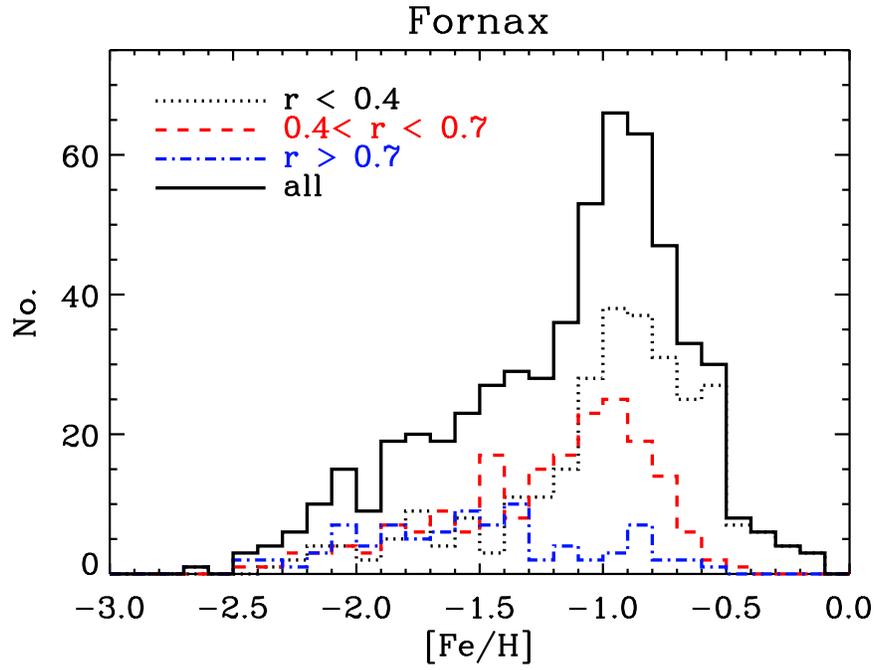}
\caption{Histogram of metallicity measurements for individual RGB stars 
in the Fornax dSph (solid line: all velocity members; dotted line: velocity members 
within $r<$ 0.4 deg from the centre; dashed line: velocity members between 
0.4$<r<$0.7 deg from the centre; dashed-dotted line: velocity members at $r>$0.7 deg 
from the centre). }
\label{fig:histo_met}
\end{center}
\end{figure}

Figure~\ref{fig:met} shows the metallicity distribution with elliptical radius for 
the Fornax members kinematically selected from our VLT/FLAMES data. 
A variation of metallicity with elliptical radius is clearly visible: a metal poor 
population ($-2.8 <$ [Fe/H] $< -1.2$) is seen throughout 
the galaxy; a 
more metal rich population ([Fe/H]$_{\rm peak}\sim -$0.9) is present out to 
0.7 deg from the centre; and an even more metal rich tail, extending out to 
[Fe/H]$=-$0.1, is present mainly in the inner 0.3-0.4 deg.
This is quantified as an histogram in Fig.~\ref{fig:histo_met}, where the sample is divided 
into 3 spatial bins (inner $r<$0.4, middle $0.4<r<0.7$, and outer $r>$0.7 ) 
and we can see that the  
metal poor component, centred at [Fe/H]$\sim-$1.7, is present 
in each bin and is thus spread throughout the galaxy; 
another component, centred at [Fe/H]$=-$1, with width at half 
peak of about 0.3, is present in the middle and inner regions; and only the innermost region 
contains the most metal rich component. 

Such a large sample of velocity and metallicity measurements, 
covering a much larger area than previous studies, 
gives us the possibility to explore the relationship 
between the kinematics and metallicity in this galaxy.  
We divided the metallicity distribution into two parts: stars more metal 
rich (MR) and more metal poor (MP) than [Fe/H]$=-$1.3. This value
of the metallicity was chosen to try and minimise the overlap between 
the two metallicity components. However a slightly different cut 
(e.g. [Fe/H]$-$1.2 or $-$1.4) does not significantly change the conclusions of our analysis. 

\begin{figure}
\begin{center}
\includegraphics[width=\textwidth]{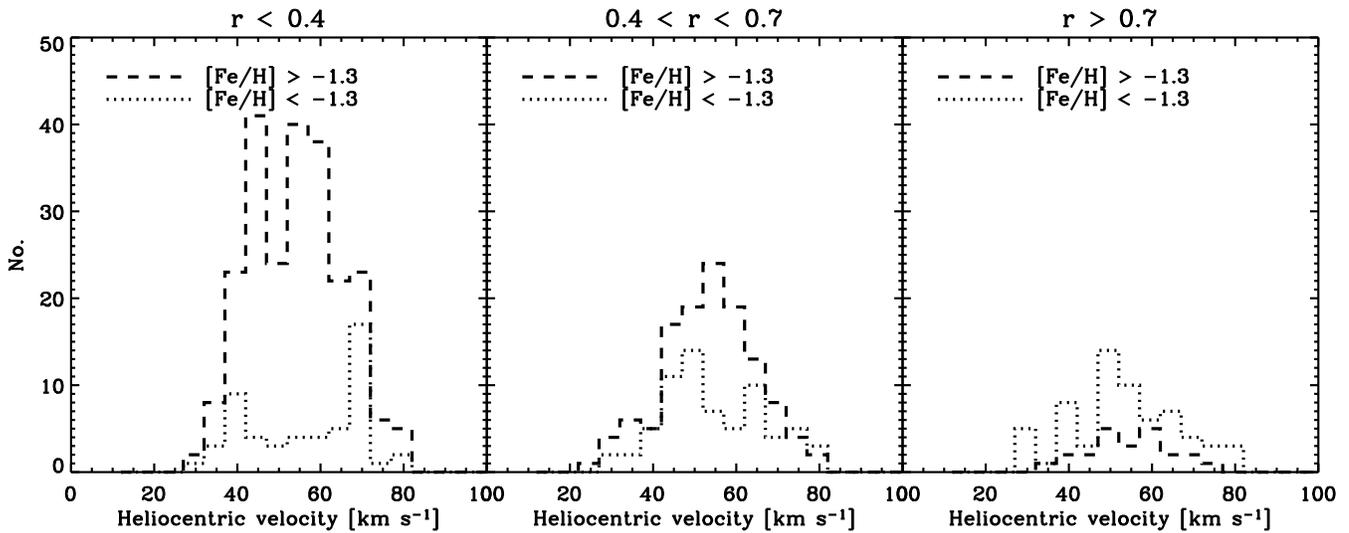}
\caption{Velocity histograms for 3 different distance bins for metal rich (dashed) and 
metal poor (dotted) stars in the Fornax dSph. The ``metal rich'' stars display a colder 
kinematics than the ``metal poor'' stars (see Table~3). Note the flat/double-peaked velocity distribution 
of MP stars in the inner bin.}
\label{fig:histo_vel_bin}
\end{center}
\end{figure}

\begin{figure}
\begin{center}
\includegraphics[width=90mm]{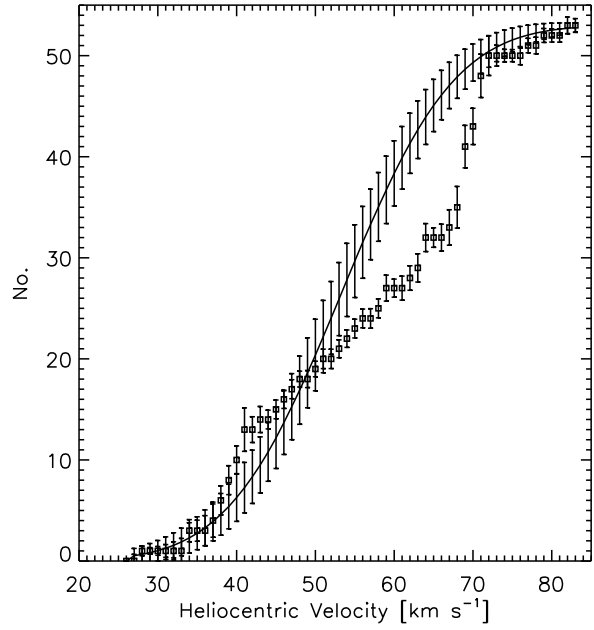}
\caption{The squares with errorbars show the cumulative velocity distribution for MP stars ([Fe/H]$<-1.3$) 
in the inner distance bin in the Fornax dSph (see Fig.~19). The dotted line is the cumulative distribution of a Gaussian with 
mean velocity and dispersion as measured for MR stars in the inner distance bin and 
the same number of stars as in the MP component within $r<$ 0.4 deg (solid line with 
errorbars). A KS-test 
gives a probability of 0.7\% for MP and MR stars in the inner bin to be drawn from the same distribution.}
\label{fig:cum_mp}
\end{center}
\end{figure}

Figure~\ref{fig:histo_vel_bin} shows the velocity distributions for the two metallicity 
components in 3 different spatial bins (inner $r<$ 0.4 deg, 
middle 0.4 $<r<$ 0.7 deg, outer $r>$ 0.7 deg). The metal poor population 
exhibits a larger velocity dispersion than the metal rich 
population\footnote{The velocity dispersion values we list are from the 
weighted standard deviation and do not include the broadening due to measurement errors. 
The true dispersion $\sigma_{\rm true}$ is approximately $\sigma^2_{\rm true}= \sigma^2_{\rm obs} - \sigma^2_{\rm meas}$, 
where $\sigma_{\rm obs}$ is the observed velocity dispersion and 
$\sigma_{\rm meas}$ is the average measured error in velocity. For our FLAMES data 
the average error in velocity is $\sim$ 2 \kms; the resulting values 
of the velocity dispersion are thus only marginally inflated, for example if 
$\sigma_{\rm obs}= 10$ \kms then $\sigma_{\rm true}=$ 9.8 \kms.} 
(see Table~\ref{tab:flames});  
furthermore,  
in the first bin the velocity distribution of the metal poor stars is far
from being Gaussian: it is flat or even double peaked (with peaks 
between 37-42 \kms and 67-72 \kms). Changing the 
binning in the histogram leaves the main features of 
the velocity distribution unchanged. 
 Figure~\ref{fig:cum_mp} shows the cumulative function of the MP stars at $r<0.4$ deg 
compared to the cumulative distribution for a Gaussian with the 
same mean velocity and dispersion as the metal rich component 
in the same distance bin. The two cumulative functions are 
very different, and this also shows that the cumulative function 
of the MP stars is similar to what 
is expected for a uniform distribution, with two peaks at $\sim$ 40 \kms 
and $\sim$ 70 \kms. 

An issue is if the velocity distribution of MP stars in the first bin 
is artificially biased by our choice of the metallicity cut, namely if by assigning 
metal poor stars to the metal rich component we could cause the small observed 
number of MP stars 
with velocity close to the systemic in the inner bin. 
This can be explored by changing the cut in 
metallicity, e.g. using
[Fe/H] $< -1.6$ for the metal poor component, and
[Fe/H] $> -1.1$ for the metal rich one. We find the same features for the MP distribution at $r<$~0.4 deg 
as in the case with a metallicity
division at [Fe/H]$=-1.3$, showing that the choice of a different metallicity cut does not alter our conclusions 
and accentuates the differences between  
the velocity distributions for metal rich and metal poor stars.

We tested to see if the differences in velocity distribution 
between the metal poor and the metal rich components are statistically 
significant: can the metal poor 
population in the inner bin be drawn from the metal rich and how significant would this be? 
A two-sided KS-test applied to the velocity distributions gives a 
probability of 0.7\%, 0.4\% and 16\% that the MP stars in the first bin might be 
drawn from the same distribution as respectively: the MR stars in the inner bin; the MR 
stars in the middle bin; the MP stars in the middle bin. 
Thus the differences in the velocity distribution of MR and MP stars 
are unlikely to be an artifact of the observed number of stars, and instead 
point to significantly different kinematic behaviour of different metallicity 
components. 

We also tested to see if the differences between the velocity distribution 
of MP stars in the inner and middle bins, reflected 
in the relatively low value of the K-S test, are due to the peculiarities of  
the velocity distribution of the MP stars at $r<$ 0.4 deg, or 
are intrinsic. Thus, we artificially ``removed'' the two velocity peaks by considering 
as the ``expected'' number of stars at the velocities 
of the peaks as the average of the number of stars in the adjacent bins. 
We then randomly removed the stars in ``excess'' 
from the velocity peaks (5.5$\pm$2.3 for the peak at $\sim$ 40 \kms 
and 14$\pm$4 at $\sim$ 70 \kms). The velocity distribution of the remaining stars 
 is compatible with a Gaussian distribution with velocity dispersion 
$\sigma= 13.4 \pm 1.74$ \kms (probability of 99.85\% from a KS-test).
In this case MP stars in the inner and middle distance bins 
have a 93.7\% probability to have been drawn from the same velocity distribution.

Thus Fornax dSph shows clear differences in the kinematics of its 
MR and MP component. Contrary to expectations, part of the metal poor component, 
arguably the oldest component, displays 
non-equilibrium kinematic behaviour at $r<$ 0.4 deg. A possible explanation for this is that Fornax 
recently captured external material 
which is disturbing the underlying distribution. Since we detect signs of 
disturbance only in the MP component, we argue that part of the object accreted by Fornax 
must have been dominated by stars more metal poor than [Fe/H]$<-1.3$. 

\subsection{Age determination}

It is well known that the main uncertainty in deriving the absolute
ages of stars in a CMD of a complex stellar population (ie. a galaxy)
is that the position of a star changes depending degenerately upon
both age and metallicity. We can break this degeneracy using
metallicities derived from spectroscopy.  In Fornax we can use our
spectroscopic metallicities of 562 RGB stars to determine which
isochrone set to use to determine the ages of these stars, and thus
produce an age-metallicity relation for the galaxy over the age range
covered by RGB stars ($>$1Gyr).  The ages we determine in this way
remain uncertain in absolute terms due to the limitations in the
stellar models used to create the isochrones, but in relative terms
the ages are accurate. Given the low systemic velocity of Fornax
there will be foreground stars contaminating our samples.

We chose to use the Yonsei-Yale isochrones \citep{yi2001, kim2002}
because they cover the range of ages and metallicities we require in a
uniform way, and they allow for a variation in [$\alpha$/Fe]. They also
provide a useful interpolation programme which allows us to
efficiently calculate the exact set of isochrones to compare with each
spectroscopic metallicity. These isochrones did have the problem that
they did not always extend up to the tip of the RGB for young metal
rich stars, but comparison with the Padua isochrones (Girardi et
al. 2000) of the same metallicity suggested that we could extrapolate
these Y-Y isochrones to the tip of the RGB which allowed us to
determine ages of the young metal rich stars in our sample.

\begin{figure}
\begin{center}
\includegraphics[width=90mm]{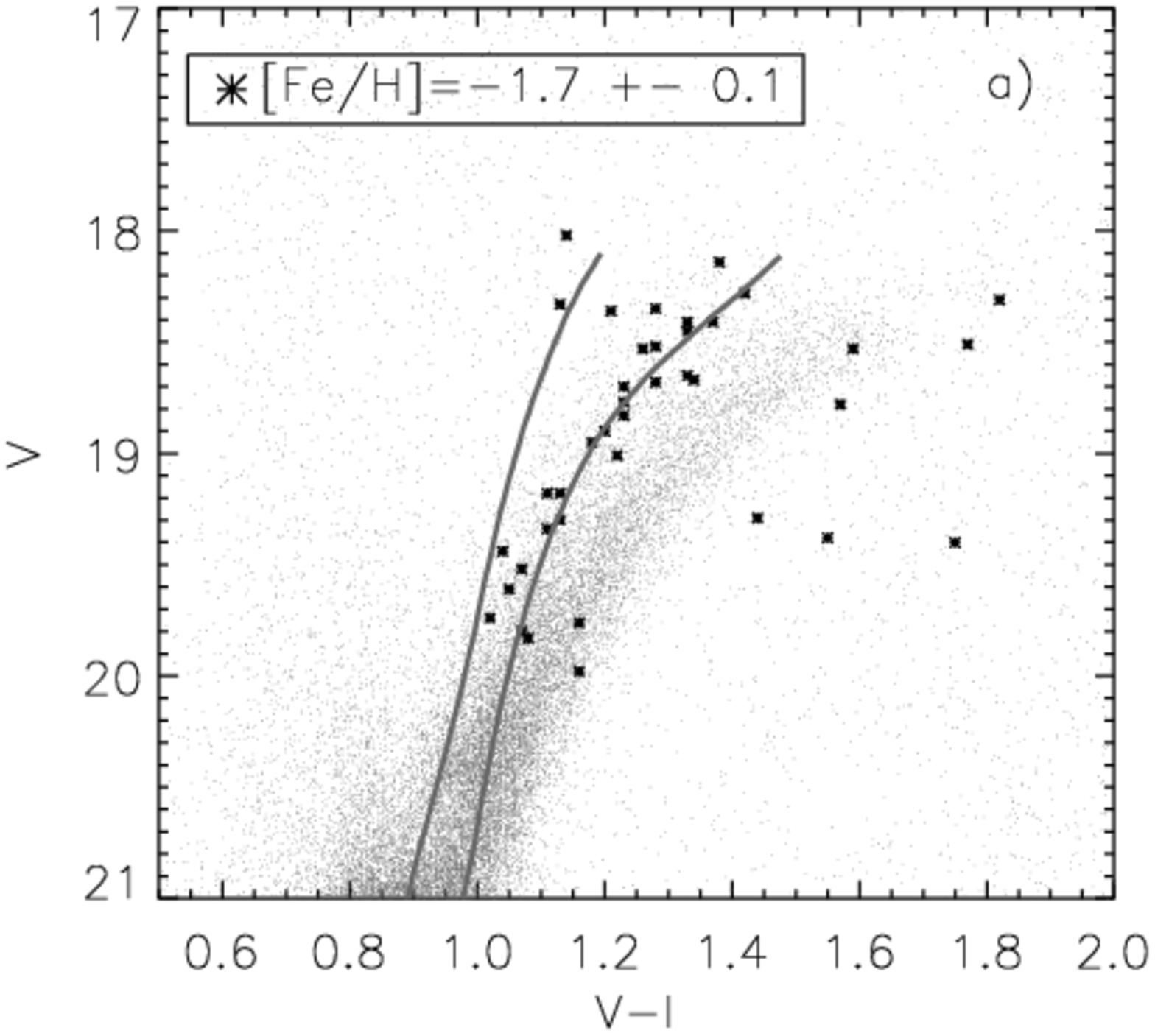}
\includegraphics[width=90mm]{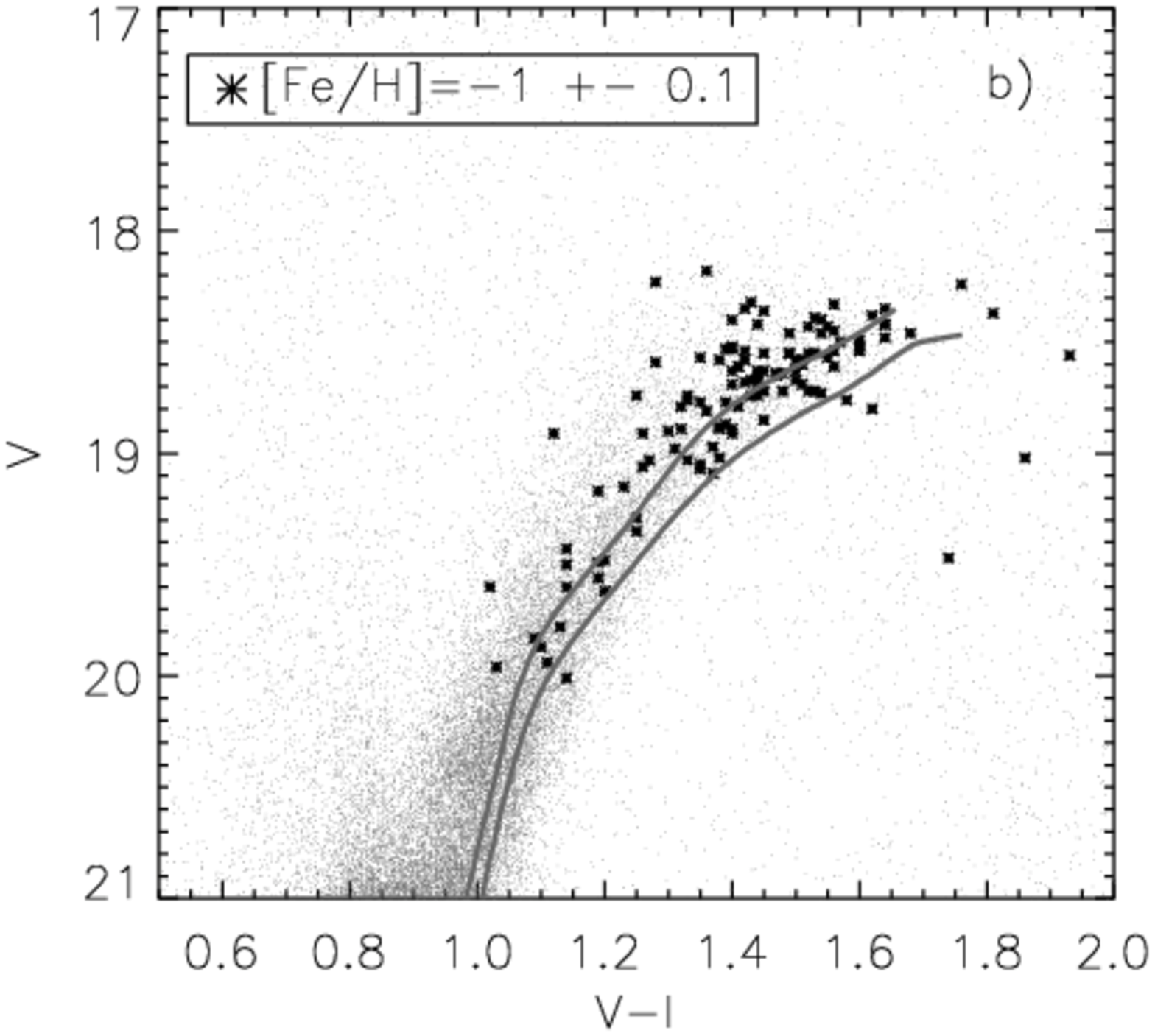}
\includegraphics[width=90mm]{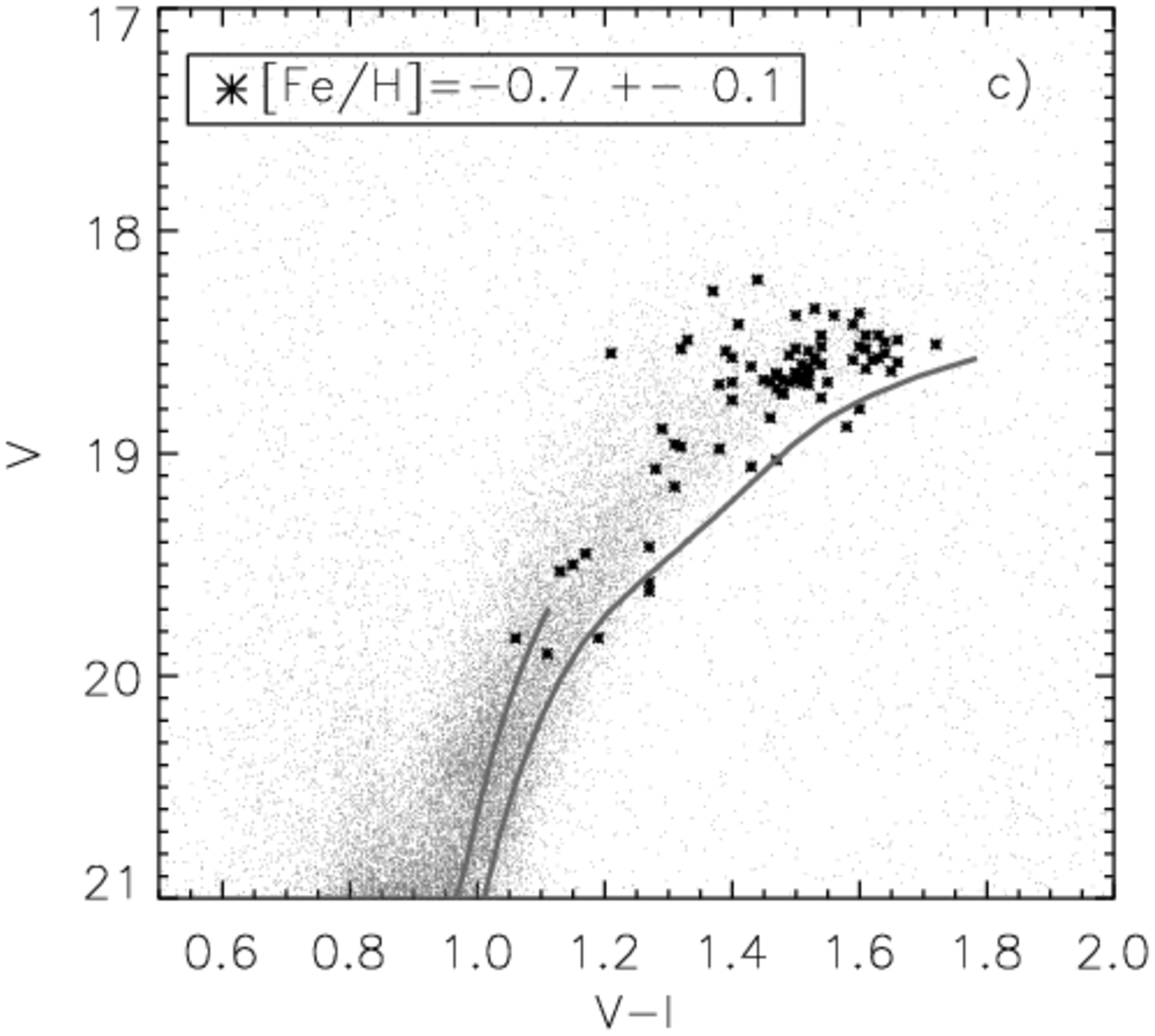}
\caption{
The comparison between measured metallicities and colours of Fornax stars (asterisks) with the 
theoretical isochrones (lines) for the metallicity of the stars. 
a) Stars with [Fe/H]$=-1.7\pm$0.1 
and two YY isochrones of [Fe/H]$=-1.7$ and age 2 and 13 Gyr (from left to right). 
b) Stars with [Fe/H]$=-1\pm0.1$ and two 
YY isochrones of [Fe/H]$=-1$ and age 2 and 5 Gyr (from left to right). 
c) Stars with [Fe/H]$=-0.7\pm$0.1 and two 
YY isochrones of [Fe/H]$=-0.7$ and age 1 and 2 Gyr (from left to right).}
\label{fig:cmd_mp_iso}
\end{center}
\end{figure}

In Fig.~\ref{fig:cmd_mp_iso}a we plot the CMD of the 39 stars in our
spectroscopic sample with [Fe/H]$=-1.7\pm 0.1$ and over-plot the Y-Y
isochrones of the same metallicity for two different ages: the
majority of metal poor stars fall on the blue side of the RGB and are
consistent with old ages ($>$10Gyr old), and thus 
can be associated with the ancient component 
from the photometric analysis in Sect.~3.  The stars found outside the
range of the isochrones to the red ($V-I\gtrsim$ 1.4) may be Galactic
contamination.  There are also several quite blue stars (at $V-I\sim$
1.1; $V\sim$ 18.3) which are consistent with having young ages (2
Gyr), and these are most likely to be foreground contamination.  The
more metal rich stars for which we have spectroscopy, with [Fe/H]$=-1\pm
0.1$ (119 stars), are shown in Fig.~\ref{fig:cmd_mp_iso}b. They are
generally consistent with young, 2-5Gyr old, isochrones and match the
dominant intermediate age component found in the photometric
analysis. Fornax members at higher metallicity ([Fe/H]$=-0.7\pm 0.1$)
are better represented by ages of 1.5 -2 Gyr (Fig.~\ref{fig:cmd_mp_iso}c).

\begin{figure}
\begin{center}
\includegraphics[width=120mm]{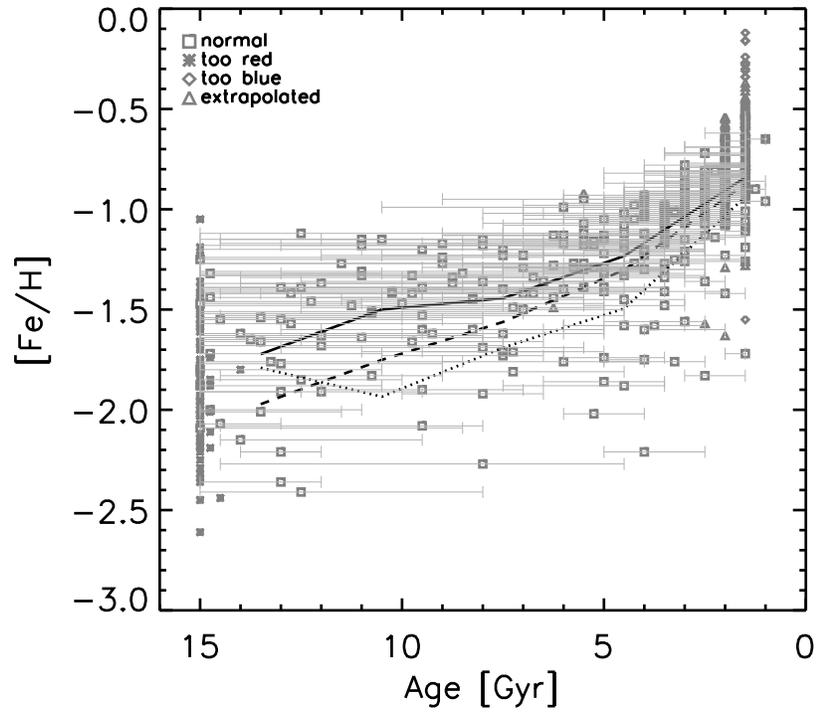}
\caption{Ages and errorbars derived from isochrones fitting of the VLT/FLAMES 
spectroscopic sample of RGB stars in Fornax dSph. The lines shows the average 
value of [Fe/H] as a function of age assuming [$\alpha$/Fe]$=0$ (solid), [$\alpha$/Fe]$=0.3$ (dotted), 
and [$\alpha$/Fe] decreasing with [Fe/H] (dashed). The stars plotted as asterisks and diamonds 
were excluded from the analysis because they fell outside the age range of the isochrones 
for the considered metallicity.}
\label{fig:ages}
\end{center}
\end{figure}

In Fig.~\ref{fig:ages} we show the age-metallicity relation obtained
for Fornax from fitting Y-Y isochrones to our entire spectroscopic
sample.  We used [$\alpha$/Fe]$=0$, in agreement with the average
value from preliminary HR measurements of RGB stars in the centre of
the Fornax dSph \citep{letarte2006b}. We derived the errors in the age
determination from the errors in magnitude and colour of the
photometry, and assuming a metallicity error of 0.1 dex for each star.
This figure contains 466 stars (out of the total sample of 562 stars),
among which 103 were young stars which were too bright for the young
isochrones (but fainter than the tip of the RGB); 97 stars fell
completely outside the age range of the isochrones and therefore had
to be excluded.

If we change the [$\alpha$/Fe] assumed, and repeat our age
determination for [$\alpha$/Fe]$=$ 0.3; and for a variation in
$\alpha$ with [Fe/H] ([$\alpha$/Fe]$= 0.2$ for [Fe/H] $<-1.7$ and
decreasing to [$\alpha$/Fe]$= 0$ between [Fe/H] $ = -1.7 \& -1.0$ and
then remaining constant), as found in HR studies of the central region
of Fornax \citep[]{letarte2006b} we obtain slightly different results
as shown by the relations plotted in Fig. ~\ref{fig:ages}.  The
different trend is enhanced by the fact that previously excluded
groups of low metallicity stars when using [$\alpha$/Fe]$>0$ isochrones
could now be included.  Thus the $\alpha$-element abundance of a star
has an impact on deriving accurate ages, and it would clearly be
desirable to be able to correct for this effect for each star.
Unfortunately [$\alpha$/Fe] determination requires HR spectroscopy and
for the large sample here this is a daunting task. However we can make
use of the HR study of the central region and use the general trends
found there.  HR studies \citep[]{letarte2006b} find that
[$\alpha$/Fe]$< 0$ for [Fe/H]$\gtrsim -1.0$, however as there are no
currently available sets of isochrones for [$\alpha$/Fe]$<0 $, it is
not clear how this will affect our age estimates. Assuming the
interpolation between [$\alpha$/Fe] = 0.3 and 0. continues to $-$0.3
this would go in the direction of reconciling the discrepancies
between the observations and the isochrones at high metallicities.

\section{Discussion}

\subsection{Age \& Metallicity Gradients}

One of the main results of our imaging and spectroscopic survey of the
Fornax dwarf spheroidal galaxy is the presence of a population
gradient. The stellar population shows a clear variation in age and
metallicity as a function of radius (e.g. Figs.~6, 14 and 19).  We can associate most of the
more metal poor component with an ancient population ($> 10$Gyr), and
most of the more metal rich component with an intermediate age
population (2-8 Gyr) with overlapping 
metallicity distributions. 

Population gradients have been detected in several other dwarf
spheroidal galaxies in the Local Group from imaging 
(e.g., Harbeck et al. 2001), and spectroscopy (e.g., Tolstoy et al. 2004).
This must point to some common process in the formation of stars in these
small systems, as this variation most likely reflects the original 
spatial distribution of gas from which the different stellar
populations were formed.  It is unlikely that the spatial distribution
of one stellar component can have changed significantly with
respect to another over time. Although weak encounters could
diffuse an older component, it is unlikely to be an important effect
because the relaxation time (Binney \& Tremaine 1987) 
of Fornax is $> 1.5 \times 10^{10}$ years,
i.e. longer than the age of the ancient population. Therefore, the
present spatial distribution of stars of different ages gives us an
indication of how the gas from which they were formed was distributed,
and requires an explanation for how gas has apparently been
progressively ``removed'' from the outer regions, and more recently,
apparently, completely from the system.

Star formation is likely to have a dramatic impact on the interstellar
medium of these small galaxies. It is possible that gas is frequently 
blown away as the result of supernovae type II
explosions of massive stars \citep[e.g.][]{maclow1999, mori2002}. 
This expelled gas should be able to fall back onto the
galaxy, if the galaxy has a large enough potential, and on its return
the gas will sink deeper into the central regions, where further
generations of star formation can occur, with potentially different
spatial, kinematic, metallicity and age characteristics.  This 
next generation stellar population will likely
be a more centrally concentrated, younger, more metal rich
component.  In this case we should expect a direct correlation between
the mass of the galaxy and the fraction of gas retained by the system,
as the deeper potential well of the most massive dSphs is more capable
of retaining more of the original gas (e.g., Ferrara \& Tolstoy 2000). 
Consistent with this picture, 
Fornax differs from the less massive Sculptor dSph 
in having a much more extended star formation history.

However this correlation is not a straight forward prediction 
of the properties of all the dwarf galaxies in the Local Group. 
This is because it is not only the mass of galaxies that determines their ability
to retain an interstellar medium; dSph typically live in a fairly active 
environment with competing effects of the tidal field of our much
larger Milky Way and several fairly large interacting companions
(e.g., the Magellanic Clouds), and the debris left around by these
interactions (e.g., the Magellanic Stream).  N-body simulations indicate that tidal stripping
combined with ram pressure stripping is a feasible combination to
explain the lack of gas at the present time in dSphs
\citep{mayer2005}.  
The importance of these
effects depends strongly on the orbital parameters of the satellite,
the apocenter-to-pericenter ratio, and the time at which the satellite
enters the Milky Way potential.
Whilst the satellite is orbiting the Galaxy, part
of its gas may be ionized by the Milky Way corona, perhaps in
combination with the cosmic UV background. The gas can be then removed
over time, with varying degrees of efficiency,
if the apocenter-to-pericenter ratio is
large enough.

Recent proper motion measurements for Fornax suggest 
a pericentric distance of the order of
the present Galactocentric distance \citep{piatek2002, dinescu2004}, 
and an almost circular orbit with an orbital period of $\sim$
4-5 Gyr \citep{dinescu2004}. Even though proper motion measurements are rather uncertain,
this would suggest that we can exclude strong interaction with the
Milky Way. Thus, as suggested by \citet{mayer2005}, an object like Fornax
may not be completely stripped of its gas, but would be able to retain
it in its central regions and thus maintain an extended star
formation history.

N-body models that predict dIrr to be the progenitors of dSphs, transformed into
spheroids by tidal stirring from the Milky Way \citep[][]{mayer2001a, mayer2001b}, 
predict an increase in the star formation activity of a 
satellite after pericentric passage.  The pericentric passage can
also provoke the formation of a bar that funnels gas into the central
region, making the gas distribution more centrally concentrated. These models are highly
dependent upon the starting conditions for the satellite. It is found
that an apocentre-to-pericentre ratio of $\sim$ 5 and a pericentre
distance of $\sim$ 50 kpc would be required to transform a dIrr into a
dSph. Fornax has never been specifically modeled, however, as it is
now thought to be close to its pericentre and the previous
passage probably occurred around 4-5 Gyr ago, this would correspond well to the time
frame of the last period of intense star formation. This scenario could
explain both the variation in the spatial distribution of the Fornax
stellar population and the large fraction of intermediate age
stars. Specific simulations, emulating the orbital parameters of
Fornax, are required for a more detailed comparison between these
observations and N-body simulations.

\subsection{Peculiar Kinematics}

In addition to the different spatial distribution of ancient,
intermediate age and young stars, we also find a different kinematic
behaviour of metal rich and metal poor stars.  This behaviour appears
similar to what is seen, for example, in our Milky
Way. The stellar bulge of our Galaxy is much less extended and
kinematically colder than the stellar halo.  This distinction may
merely reflect the different spatial distribution required by a star
moving with a larger velocity to reach larger distances or it may be 
telling us something more fundamental about the different conditions of 
the formation of these two different components. To test this
we can use the Jeans equation to derive the line of sight velocity
dispersion predicted for stars following the different density
profiles of old and intermediate age stars \citep{binney1987}. This
calculation shows that it is difficult to interpret the observed
differences in velocity dispersion of MR and MP stars as being only
due to their different spatial distribution \citep{Battaglia2006a},
and the same result is found in N-body simulations
\citep[e.g][]{kawata2006}. In a future paper we will present detailed
kinematic modelling of these distinct stellar populations, taking into
account the possibility of different orbital properties for the
two populations \citep{Battaglia2006a}.

Another aspect that deserves further investigation is the apparently
{\it non-equilibrium}
 kinematics of the metal poor, presumably the oldest,
stellar population in the center of Fornax. A possible interpretation of
this is that we are seeing a remnant of accreted material. Given the
presence of 5 globular clusters in Fornax, it would not be surprising for
this to be a disrupted globular cluster. In this case we would expect
the metallicity distribution of the stars forming the velocity feature
to be extremely narrow, however the metallicity of the stars in the
double peak ranges from $-2\lesssim$[Fe/H]$\lesssim -1.4$, showing a
much larger spread than can be accounted for by a globular cluster.

An alternative explanation, as proposed by \citet{coleman2004,
coleman2005}, is that Fornax accreted a smaller galaxy at some point in
the recent past.  These authors propose the accretion of a gas-rich
dwarf $\sim$ 2 Gyr ago.  In this case the MP stars with unrelaxed
kinematics could be the remnant of the disrupted old stellar component of
this accreted galaxy, and the asymmetric spatial distribution of the
young stars (BL and MS) could be a result of the gas which subsequently formed 
stars as a consequence of the accretion process.
Measuring the metallicities of the young stars in Fornax could help to answer this
question by determining whether or not they are consistent with the bulk of the
older stellar population in Fornax. Unfortunately we do not have 
spectroscopic metallicities of these stars, but
the indications from the colour of the Blue~Loops in the CMD (see
section 3.3.3) suggest that these stars have a metallicity of
[Fe/H]$\sim -0.85 \pm 0.15$, consistent with the mean of the youngest 
RGB stars (age $>$1-2 Gyr) in our VLT/FLAMES sample.  
However, our FLAMES sample also contains a significant number of stars
with higher metallicity, up to [Fe/H]$\sim-0.1$.  Assuming we can
directly compare metallicities derived with these two different
methods, this could be an indication that Fornax has recently
($<$1-2 Gyr ago) been polluted with external gas of lower metallicity than 
the gas in Fornax at that time. This
would also explain the peculiar kinematics of the MP stars in the
centre of Fornax.  However, this conclusion requires further
confirmation, such as high resolution follow-up of the metal rich 
RGB stars in the central region (Letarte et al. 2006b), 
spectroscopic analysis of the Blue~Loop and/or main sequence stars, and detailed dynamical modelling.

\section{Summary}
We have presented results from accurate ESO/WFI photometry 
of resolved stars in the Fornax dSph galaxy covering the 
entire extent out to its nominal tidal radius, and 
also velocity and metallicity measurements from our VLT/FLAMES spectroscopy 
of 562 RGB stars also out to the nominal tidal radius. 

Using our ESO/WFI photometry 
we re-derive basic structural parameters of Fornax such as central position, 
ellipticity and position angle.
In common with previous imaging studies we show that 
the Fornax dwarf spheroidal galaxy
contains 3 stellar components: an ancient component, with ages between
10-15 Gyr old, which is spatially extended; an intermediate age component,
with ages 2-8 Gyr old, that contains the bulk of the stellar population
($\sim$ 40\%), which is more centrally concentrated and
less extended with respect to the older component; a young population,
with stars younger than $\sim$ 1 Gyr old. The youngest component 
($<$1 Gyr old) is the most
centrally concentrated and is completely absent beyond a radius of 0.4 deg 
from the centre, and
it shows an asymmetric, disc-like (or bar-like) distribution.  Within
this component, the 200-300 Myr old stars show the most asymmetric
and centrally concentrated distribution. The colour of the plume formed in the CMD by Blue~Loop
stars indicates that this young population has a metallicity [Fe/H]
between $-0.7$ and $-1.0$.  Spatial analysis shows that
substructures might be present in the ancient stellar
population represented by the B-RGB.

The VLT/FLAMES spectroscopy in the CaT wavelength region for a
sample of 562 Fornax RGB velocity members shows that the mean metallicity
of the stellar population changes with radius with the central regions 
more metal rich than the outer regions. No stars more metal poor than
[Fe/H]$\lesssim -2.7$ are found anywhere in the galaxy, but we can distinguish a metal poor
component, with a mean of [Fe/H]$\sim -1.7$ and extending from
$-2.5 \lesssim$[Fe/H]$\lesssim -1$, that is found throughout the 
galaxy; a more metal rich component, with a mean of [Fe/H]$\sim -1$ and
extending from $-1.5 \lesssim$[Fe/H]$\lesssim -0.5$, present mainly
within 0.7 deg from the centre and is thus more spatially concentrated than
the metal poor stars; a metal rich tail, [Fe/H]$\lesssim-0.1$, is 
found at radii less than 0.4 deg from the centre. ``Metal rich'' and ``metal poor''
populations (stars with [Fe/H]$>-1.3$ and $<-1.3$, respectively) show
different kinematics: metal rich stars have a colder velocity
dispersion than metal poor stars, and these differences are not dependent
on the metallicity cut. The metal
poor component shows signs of non-equilibrium kinematics in the inner
regions ($r<0.4$ deg), arguably due to the accretion of external
material. A more complete modelling of chemo-dynamics of Fornax is the subject 
of a future paper (Battaglia et al. 2006a).

\begin{acknowledgements}
GB would like to thank Andrew Cole for useful suggestions. 
ET gratefully acknowledges support from a
fellowship of the Royal Netherlands Academy of Arts 
and Sciences 
AH acknowledges financial support from the Netherlands
Organization for Scientific Research (NWO). 
MDS would like to thank the NSF for partial support under AST-0306884. 
KAV would like to thank the NSF for support through a CAREER
award, AST 99-84073, as well as the University of Victoria
for additional research funds.
\end{acknowledgements}

\bibliographystyle{aa}
\bibliography{fnx_biblio}

\clearpage


\begin{longtable}{lccccc}
\caption{\label{tab:photo_obs} Table of Fornax ESO/WFI observations. 
The seeing is the average stellar FWHM from the final image.}\\
\hline
\hline 
Name &  Filter & UT of observation &   airmass &   exptime &  seeing (arcsec) \\
\hline
\endfirsthead
\caption{continued.}\\
\hline
\hline 
Name &  Filter & UT of observation &   airmass &   exptime &  seeing (arcsec) \\
\hline
\endhead
\hline
\endfoot
FNX-1&   V & 07-Jan-2005 02:13 &      1.1  &    3x300s &     0.87 \\
     &   I &       &    1.1  &    3x300s &     0.89 \\
FNX-2&   V & 07-Jan-2005 02:55 &      1.2  &    3x300s &     0.90 \\
 &   I &    &       1.2  &    3x300s &     0.80 \\
FNX-3&   V & 07-Jan-2005 03:39 &      1.3  &    3x300s &     0.93 \\
 &   I &               &     1.3  &    3x300s &     0.93 \\
FNX-4&   V & 07-Jan-2005 04:22  &     1.5  &    3x300s &     1.13 \\
 &          I &              &    1.6  &    3x300s &     0.87 \\
FNX-13&    V & 08-Jan-2005 02:19  &     1.1  &    3x300s &     0.75 \\
 &          I &       &    1.1  &    3x300s &     1.19 \\
FNX-5&   V &  08-Jan-2005 02:55  &     1.2  &    3x300s &     0.82 \\
 &   I &              &    1.2  &    3x300s &     1.12 \\
FNX-6&   V &     08-Jan-2005 03:26 &      1.3 &     3x300s &     1.03 \\
 &   I &        &   1.3 &     3x300s &     1.00 \\

FNX-8&     V & 01-Feb-2005 01:54  &     1.3 &     3x300s &     0.98 \\
 &   I &       &    1.3 &     3x300s &     0.94 \\
FNX-9&     V & 01-Feb-2005 02:18 &      1.4  &    3x300s &     1.09 \\
 &   I &      &     1.5 &     3x300s &     0.97 \\
FNX-14&   V & 01-Feb-2005 02:54 &      1.6 &     3x300s &     1.01 \\
 &   I &      &     1.7 &     3x300s &     0.93 \\
FNX-16&   V & 02-Feb-2005 01:35 &      1.2 &     3x300s &     1.09 \\
 &   I &        &     1.3 &     3x300s &     1.21 \\
FNX-7&   V & 01-Feb-2005 01:16 &      1.2 &     3x300s &     1.02 \\
 &   I &      &     1.2 &     3x300s &     0.97 \\
FNX-21&   V & 03-Feb-2005 02:10 &      1.4 &     3x300s &     1.02 \\
 &   I &      &     1.5 &     3x300s &     1.09 \\

SA98 &     V &     06-Jan-2005 05:08  &     1.2  &   10s,30s &   0.92,0.95 \\
   &        I &                 05:19  &     1.2 &    10s,30s &   0.93,0.94 \\
SA95 &     V &     07-Jan-2005 01:32   &    1.2  &   10s,30s &   1.06,1.03 \\
   &        I &                 01:44  &     1.2 &    10s,30s &   1.06,0.80 \\
RU-149 &   V &     07-Jan-2005 05:02 &      1.2 &     5s,30s &   0.89,0.95 \\
       &   I &                 05:12   &    1.2   &  10s,30s &   1.66,1.16 \\
SA95 &     I &     09-Jan-2005 02:48  &     1.2  &   10s,30s &   1.16,1.26 \\                      
RU-149 &   V &     31-Jan-2005 00:46  &     1.0  &    5s,30s &   3.55,3.61   \\   
  &         I &                 01:00  &     1.0  &   10s,30s &   2.70,2.56 \\
SA98 &     V &     01-Feb-2005 00:30   &    1.4  &   10s,30s &   0.98,0.93 \\
  &         I &                 00:42  &     1.3  &   10s,30s &   1.14,1.24 \\
SA95 &     V &     02-Feb-2005 00:44  &     1.2  &   10s,30s &   1.24,1.22 \\
   &        I &                 00:58  &     1.2  &   10s,30s &   1.05,1.02 \\
SA101 &    V &     02-Feb-2005 08:33  &     1.5  &   10s,30s &   0.94,1.15 \\
   &        I &                 08:41  &     1.5  &   10s,30s &   1.30,1.22 \\
PG1323-086&V &     02-Feb-2005 09:10  &     1.1  &    5s,30s &   0.86,0.91 \\
   &        I &                 09:22  &     1.1  &   10s,30s &   1.20,1.07 \\
SA95 &     V &     03-Feb-2005 01:37  &     1.3   &  10s,30s &   0.95,0.87 \\
   &        I &                 01:48  &     1.3  &   10s,30s &   0.91,0.80 \\
SA107&     V &     03-Feb-2005 09:17  &     1.4   &  10s,30s &   0.79,0.88 \\
   &        I &                 09:28  &     1.3  &   10s,30s &   0.69,0.74 \\
SA98 &     V &     04-Feb-2005 01:02  &     1.2   &  10s,30s &   1.16,1.23 \\
   &        I &                 01:13  &     1.2  &   10s,30s &   0.63,0.74 \\
\hline
\end{longtable}

\clearpage

\begin{table}
\caption{Table of Fornax parameters. 1. This work. 2. Mateo et al. (1998). 
The position angle is defined as the angle between North and the major-axis of the galaxy 
measured counter-clockwise; the ellipticity is $e= 1 - b/a$. }
\label{tab:par}
\centering
\begin{tabular}{lcc}
\hline
\hline 
$\alpha_{\rm 2000}$ & $2^h 39^m 52^s$ & 1 \\
$\delta_{\rm 2000}$ & $-34^{\circ} 30' 49''$ & 1 \\
Position angle  & $46.8^{\circ} \pm 1.6^{\circ}$ & 1 \\
Ellipticity & $0.30 \pm 0.01$ & 1 \\
$L_V$ & $1.55 \times 10^7 L_{\odot}$& 2 \\
Distance &  $138 \pm 8$ kpc & 2 \\
$V_{\rm h}$ & $54.1 \pm 0.5$ \kms & 1 \\
${\rm [Fe/H]}_{\rm mean}$ & $-1.15$ & 1 \\
\hline
\end{tabular}
\end{table}

\begin{table} 
\caption{Parameters of best-fitting King model (core radius, $r_{\rm c}$, tidal radius, $r_{\rm t}$), 
Sersic model (Sersic radius, $R_{\rm S}$, shape parameter, $m$), exponential model (scale radius, $r_{\rm s}$) 
and Plummer model (scale radius, $b$) 
for all stellar populations, intermediate age and old stellar components in Fornax dSph. For each model 
we also compute the half-number radius, $r_{\rm h}$; for the Plummer model this coincides with $b$. The 
errors in the parameters are formal errors from the fitting procedure.
}
\label{tab:par_oldint}
\centering
\begin{tabular}{lccccccccc}
\hline
\hline
\multicolumn{1}{l}{{}} &  \multicolumn{3}{c}{{King}} & \multicolumn{3}{c}{{Sersic}} &
\multicolumn{2}{c}{{Exponential}} & \multicolumn{1}{c}{{Plummer}} \\
\hline
  & $r_{\rm c}[']$  & $r_{\rm t}[']$ & $r_{\rm K,h}[']$ & $R_{\rm S}[']$  & $m$ & $r_{\rm S,h}[']$ & 
$r_{\rm s}[']$ & $r_{\rm Ex,h}[']$ & $b[']$ \\
\hline
All            &   17.6$\pm$0.2 & 69.1$\pm$0.4 & 19.1 & 17.3$\pm$0.2 & 0.71 $\pm$ 0.01 & 18.6 &
 11.0$\pm$0.1 & 18.5 & 19.6$\pm$0.1\\
Intermediate   &   15.7$\pm$0.2 & 68.5$\pm$0.4 & 18.1 & 16.7$\pm$0.2 & 0.69 $\pm$ 0.01 & 17.5 &
 10.0$\pm$0.1 & 16.7 & 17.1$\pm$0.1\\
Old            &   26.8$\pm$1.1 & 77.2$\pm$1.6 & 23.7 & 24.6$\pm$0.8 & 0.62 $\pm$ 0.02 & 23.5 & 
 13.7$\pm$0.2 & 23.0 & 24.6$\pm$0.4\\
\hline
\end{tabular}
\end{table}




\begin{table} 
\caption{Data for the observed VLT/FLAMES targets in Fornax dSph. Column (1) ID; (2) $\alpha_{\rm J2000}$; (3) $\delta_{\rm J2000}$; 
(4) $V$ magnitude; (5) $\sigma_V$; (6) $I$ magnitude; (7) $\sigma_I$; (8) signal-to-noise ratio per \AA\ ; (9) [Fe/H]; (10) $\sigma_{\rm [Fe/H]}$; 
(11) membership ID (Y= kinematic member; N= kinematic non-member). 
The errors in metallicity are based on S/N computations and adjusted to agree with repeated measurements.   
} 
\label{tab:data_spectro}
\centering
\begin{tabular}{lcccccccccccccc}
\hline
\hline
ID &  $\alpha_{\rm J2000}$ &  $\delta_{\rm J2000}$ & $V$ & 
$\sigma_V$ & $I$ & $\sigma_I$ &
S/N per \AA\ & [Fe/H] &  $\sigma_{\rm [Fe/H]}$ & Member \\
\hline
\hline
\end{tabular}
\end{table}

\begin{table} 
\caption{Number of stars, weighted mean heliocentric velocity, and line of sight velocity dispersion 
for Fornax members (all mem), metal rich (MR) and metal poor (MP) 
stars at every elliptical radius (all $r$), and in 3 distance bins ($r<0.4$, $0.4<r<0.7$, $r>0.7$ deg). 
} 
\label{tab:flames}
\centering
\begin{tabular}{lccccccccc}
\hline
\hline
\multicolumn{2}{l}{{}} &  \multicolumn{2}{c}{all $r$} &  \multicolumn{2}{c}{{$r<0.4^{\circ}$ }} & \multicolumn{2}{c}{{$0.4^{\circ}<r<0.7^{\circ}$}} &
\multicolumn{2}{c}{{$r>0.7^{\circ}$ }} \\
\hline
  & all mem  &  MR & MP&  MR &  MP & MR & MP & MR  & MP          \\
\hline
number & 562 & 377 & 185 & 232 &  53 & 122   & 68 & 23  & 64 \\
$<V>$  &  54.1$\pm$   0.5    &  53.9$\pm$0.6 &  54.5$\pm$1.0 & 53.3$\pm$   0.8      & 57.7$\pm$   1.9    &  54.5$\pm$   0.9    &  53.1$\pm$   1.5 
   &  54.8$\pm$   1.9   &  53.0$\pm$
   1.7   \\
$\sigma$  &  11.4$\pm$   0.4    &  10.6$\pm$0.4  & 13.0$\pm$0.7 & 11.3$\pm$   0.5    &  13.6$\pm$   1.4    &   9.9$\pm$   0.7    &  12.3$\pm$   1.1 
   &   8.6$\pm$   1.4   &  13.1$\pm$
   1.2   \\
\hline
\end{tabular}
\end{table}

\clearpage

\clearpage

\label{lastpage}

\end{document}